\documentclass[11pt]{article}
\usepackage{amsmath,amsfonts,amssymb}
\usepackage{cite,url}
\usepackage{graphicx}
\usepackage[utf8]{inputenc} 
\usepackage[bottom]{footmisc}
\pdfoutput=1

\allowdisplaybreaks

\def\nonSM{{\rm heavy}}
\def\smallA{{\scriptscriptstyle A}}

\def\smallZ{{\scriptscriptstyle Z}}

\def\smallH{{\scriptscriptstyle H}}

\def\smallL{{\scriptscriptstyle L}}
\def\smallR{{\scriptscriptstyle R}}

\def\MS{M_S}
\def\MZ{m_\smallZ}

\def\ma{m_\smallA}

\def\beq{\begin{equation}}
\def\eeq{\end{equation}}
\def\bea{\begin{eqnarray}}
\def\eea{\end{eqnarray}}
\def\nn{\nonumber}
\def\wt{\widetilde}

\def\FH{{\tt FeynHiggs}}

\def\SHD{{\tt SusyHD}}
\def\hssusy{{\tt HSSUSY}}

\def\sq2{\sqrt{2}}
\def\drbar{{\ensuremath{ \overline{\rm DR}}}}
\def\msbar{\overline{\rm MS}}

\def\tb{\tan\beta}
\def\gl{\tilde{g}}
\def\mg{m_{\gl}}

\def\mh{m_h}
\def\mhmax{m_h^{{\rm{\scriptscriptstyle max}}}}
\def\dmhq{\Delta m_h^2}
\def\omtms{{\cal O}(\mt^2/\MS^2)}
\def\ovms{{\cal O}(v^2/\MS^2)}
\def\qew{Q_{{\rm{\scriptscriptstyle EW}}}}
\def\xq{x_{\scriptscriptstyle Q}}
\def\xu{x_{\scriptscriptstyle U}}

\def\at{\alpha_t}

\def\as{\alpha_s}

\def\oat{{\cal O}(\at)}

\def\oatas{{\cal O}(\at\as)}

\def\oatq{{\cal O}(\at^2)}

\def\mt{m_t}

\def\ti{m_{\tilde{t}_i}^2}

\def\mb{m_b}

\def\sfr{{\tilde{f}_\smallL}}
\def\sfl{{\tilde{f}_\smallR}}
\def\mfr{m_\sfl}
\def\mfl{m_\sfr}
\def\sqr{{\tilde{q}_\smallL}}
\def\sql{{\tilde{q}_\smallR}}
\def\mqr{m_\sql}
\def\mql{m_\sqr}

\long\def\symbolfootnote[#1]#2{\begingroup%
\def\thefootnote{\fnsymbol{footnote}}\footnote[#1]{#2}\endgroup}

\newenvironment{Appendix}
 {
  \setcounter{section}{0}
  \setcounter{equation}{0}
  
 }{}

\begin{document}

\begin{titlepage}

\begin{flushright}
DESY 17-033
\end{flushright}

\vspace{1cm}
\begin{center}

\vspace{1cm}

{\LARGE \bf Improved determination of the Higgs mass} 
\vskip 0.3cm
{\LARGE \bf in the MSSM with heavy superpartners}

\vspace{1cm}

{\Large Emanuele Bagnaschi,$^{\!\!\!\,a}$~ 
Javier~Pardo~Vega$^{\,b,c}$ and Pietro~Slavich$^{\,d,e}$}

\vspace*{5mm}

{\sl ${}^a$ Deutsches
Elektronen-Synchrotron (DESY), 22607 Hamburg, Germany}
\vspace*{2mm}\\{\sl ${}^b$ Abdus Salam International Centre for 
Theoretical Physics, 

Strada Costiera 11, 34151, Trieste, Italy}
\vspace*{2mm}\\{\sl ${}^c$ SISSA International School for 
Advanced Studies and INFN Trieste, 

Via Bonomea 265, 34136, Trieste, Italy}
\vspace*{2mm}\\{\sl ${}^d$ LPTHE, UPMC Univ.~Paris 06,
  Sorbonne Universit\'es, 

4 Place Jussieu, F-75252 Paris, France}
\vspace*{2mm}\\
{\sl ${}^e$ LPTHE, CNRS, 4 Place Jussieu, F-75252 Paris, France }
\end{center}
\symbolfootnote[0]{{\tt e-mail:}}
\symbolfootnote[0]{{\tt emanuele.bagnaschi@desy.de}}
\symbolfootnote[0]{{\tt jpardovega@gmail.com}}
\symbolfootnote[0]{{\tt slavich@lpthe.jussieu.fr}}

\vspace{0.7cm}

\abstract{We present several advances in the effective field theory
  calculation of the Higgs mass in MSSM scenarios with heavy
  superparticles. In particular, we compute the dominant two-loop
  threshold corrections to the quartic Higgs coupling for generic
  values of the relevant SUSY-breaking parameters, including all
  contributions controlled by the strong gauge coupling and by the
  third-family Yukawa couplings. We also study the effects of a
  representative subset of dimension-six operators in the effective
  theory valid below the SUSY scale. Our results will allow for an
  improved determination of the Higgs mass and of the associated
  theoretical uncertainty.}

\vfill

\end{titlepage}


\setcounter{footnote}{0}

\section{Introduction}

\label{sec:intro}

At the price of doubling the particle content of the Standard Model
(SM), supersymmetry (SUSY) provides elegant solutions to several open
issues, including the stability of the electroweak (EW) scale, the
nature of dark matter and the possibility of embedding the SM in a
grand-unified gauge theory. Common features of supersymmetric
extensions of the SM are an extended Higgs sector and the existence of
tree-level relations between the quartic Higgs couplings and the other
couplings of the considered model, which translate into predictions
for the Higgs boson masses. When radiative corrections are included,
those predictions are sensitive to the whole particle spectrum of the
model, and can be used to constrain its parameter space even before
the discovery of SUSY particles.

In the minimal SUSY extension of the SM, or MSSM, the mass $\mh$ of
the lightest Higgs scalar is bounded at tree level from above by
$\MZ |\cos 2\beta|$, where $\MZ$ is the $Z$-boson mass and
$\tan\beta\equiv v_2/v_1$ is the ratio of the vacuum expectation
values (vevs) of the two Higgs doublets that participate in the
breaking of the EW symmetry.  However, as has been
known~\cite{Okada:1990vk, Okada:1990gg, Ellis:1990nz, Ellis:1991zd,
  Brignole:1991pq, Haber:1990aw} since the early 1990s, the tree-level
upper bound on $\mh$ can be significantly raised by radiative
corrections involving top quarks and their SUSY partners, the stop
squarks. By now, the computation of radiative corrections to the MSSM
Higgs masses\,\footnote{\,We focus here on the MSSM with real
  parameters. Significant efforts have also been devoted to the
  Higgs-mass calculation in the presence of CP-violating phases, as
  well as in non-minimal SUSY extensions of the SM.}  is quite
advanced: full one-loop corrections~\cite{Chankowski:1991md,
  Chankowski:1992er, Brignole:1991wp, Brignole:1992uf,
  Dabelstein:1994hb, Pierce:1996zz,Frank:2006yh} and two-loop
corrections in the limit of vanishing external
momentum~\cite{Hempfling:1993qq, Heinemeyer:1998jw, Heinemeyer:1998kz,
  Heinemeyer:1998np, Heinemeyer:2004xw, Zhang:1998bm, Espinosa:1999zm,
  Espinosa:2000df, Degrassi:2001yf, Brignole:2001jy, Brignole:2002bz,
  Dedes:2003km, Martin:2002iu, Martin:2002wn} are available, and the
dominant momentum-dependent two-loop corrections~\cite{Martin:2004kr,
  Borowka:2014wla, Degrassi:2014pfa} as well as the dominant
three-loop corrections~\cite{Harlander:2008ju, Kant:2010tf} have also
been obtained. Over the years, many of the known corrections have been
implemented in widely-used codes for the determination of the MSSM
mass spectrum. In particular, \FH~\cite{Heinemeyer:1998yj} includes
full one-loop corrections to the Higgs masses from
ref.~\cite{Frank:2006yh} and dominant two-loop corrections in the
on-shell (OS) renormalization scheme from
refs.~\cite{Heinemeyer:1998np, Degrassi:2001yf, Brignole:2001jy,
  Brignole:2002bz, Dedes:2003km, Borowka:2014wla}, whereas {\tt
  SoftSusy}~\cite{Allanach:2001kg,Allanach:2014nba}, {\tt
  SuSpect}~\cite{Djouadi:2002ze} and {\tt SPheno}~\cite{Porod:2003um,
  Porod:2011nf} include full one-loop corrections to the Higgs masses
from ref.~\cite{Pierce:1996zz} and dominant two-loop corrections in
the $\drbar$ scheme from refs.~\cite{Degrassi:2001yf, Brignole:2001jy,
  Brignole:2002bz, Dedes:2003km,Dedes:2002dy}.

For the MSSM, both the discovery in 2012~\cite{Aad:2012tfa,
  Chatrchyan:2012xdj} of a SM-like Higgs boson with mass about
125~GeV~\cite{Aad:2015zhl} and the negative results of the searches
for stop squarks at the LHC~\cite{ATLAS:2016ljb, ATLAS:2016xcm,
  ATLAS:2016jaa, CMS:2016kcq, CMS:2016vew, CMS:2016hxa} favor
scenarios with a SUSY mass scale $\MS$ in the TeV range. In
particular, the observed value of the Higgs mass requires the
radiative correction to the squared-mass parameter, $\dmhq\,$, to be
at least as large as its tree-level value: if the stops are heavy
enough, this can be realized via the dominant top/stop contributions,
which are controlled by the top Yukawa coupling,
$g_t \sim {\cal O}(1)$, and are enhanced by logarithms of the ratio
between the stop and top masses. A further increase in $\dmhq$ can be
obtained if the left-right stop mixing parameter $X_t$ is about twice
the average stop mass. Roughly speaking, for $\tb$ large enough to
almost saturate the tree-level bound on the lightest-scalar mass,
$\mh\approx 125$~GeV requires the average stop mass to be somewhere
around $1$~TeV for the ``maximal'' (i.e., most favorable) value of
$X_t$, and above $10$~TeV for vanishing $X_t$. However, when the SUSY
scale is significantly larger than the EW scale, fixed-order
calculations of $\mh$ such as the ones implemented in the codes
mentioned above may become inadequate, because radiative corrections
of order $n$ in the loop expansion contain terms enhanced by as much
as $\ln^n(\MS/\mt)$ -- where we take the top mass as a proxy for the
EW scale. Indeed, a possible symptom of such heavy-SUSY malaise is the
fact that, in scenarios with TeV-scale stop masses and large stop
mixing, the spread in the predictions of those codes for $\mh$ exceeds
the theoretical accuracy of their (largely equivalent) two-loop
calculations, which was estimated in the early 2000s to be about
$3$~GeV~\cite{Degrassi:2002fi, Allanach:2004rh} in what were then
considered natural regions of the MSSM parameter space.

In the presence of a significant hierarchy between the SUSY scale and
the EW scale, the computation of the Higgs mass needs to be
reorganized in an effective field theory (EFT) approach: the heavy
particles are integrated out at the scale $\MS$, where they only
affect the matching conditions for the couplings of the EFT valid
below $\MS$; the appropriate renormalization group equations (RGEs)
are then used to evolve those couplings between the SUSY scale and the
EW scale, where the running couplings are related to physical
observables such as the Higgs boson mass and the masses of gauge
bosons and fermions.  In this approach, the computation is free of
large logarithmic terms both at the SUSY scale and at the EW scale,
while the effect of those terms is accounted for to all orders in the
loop expansion by the evolution of the couplings between the two
scales. More precisely, large corrections can be resummed to the
(next-to)$^n$-leading-logarithmic (N$^n$LL) order by means of $n$-loop
calculations at the SUSY and EW scales combined with $(n\!+\!1)$-loop
RGEs.  On the other hand, the common procedure of matching the MSSM to
a renormalizable EFT -- such as the plain SM -- in the unbroken phase
of the EW symmetry amounts to neglecting corrections suppressed by
powers of $v^2/\MS^2\,$, where we denote by $v$ the vev of a SM-like
Higgs scalar.  Those corrections can in fact be mapped to the effect
of non-renormalizable, higher-dimensional operators in the EFT
Lagrangian.

The EFT approach to the computation of the MSSM Higgs mass dates back
to the early 1990s~\cite{Barbieri:1990ja, Espinosa:1991fc,
  Casas:1994us}. Over the years, it has also been exploited to
determine analytically the coefficients of the logarithmic terms in
$\dmhq$ at one~\cite{Haber:1993an}, two~\cite{Carena:1995bx,
  Carena:1995wu, Haber:1996fp, Carena:2000dp} and even
three~\cite{Degrassi:2002fi, Martin:2007pg} loops, by solving
perturbatively the appropriate systems of boundary conditions and
RGEs. However, when the focus was on ``natural'' scenarios with SUSY
masses of a few hundred GeV, the omission of $\ovms$ terms limited the
accuracy of the EFT approach, and the effect of the resummation of
logarithmic corrections was not deemed important enough to justify
abandoning the fixed-order calculations of the Higgs mass in favor of
a complicated EFT set-up with higher-dimensional
operators.\footnote{See, however, ref.~\cite{Espinosa:2001mm} for the
  effect of dimension-six operators in a scenario with only one light
  stop.} More recently, an interest in ``unnatural'' scenarios such as
split SUSY~\cite{ArkaniHamed:2004fb, Giudice:2004tc} and high-scale
SUSY~(see, e.g., ref.~\cite{Hall:2009nd}), and then the LHC results
pushing the expectations for the SUSY scale into the TeV range, have
brought the EFT approach back into fashion. On the one hand, in
ref.~\cite{Hahn:2013ria} the authors of \FH\ combined the fixed-order
calculation of $\mh$ implemented in their code with a resummation of
the LL and NLL terms controlled exclusively by $g_t$ and by the strong
gauge coupling $g_3$. On the other hand, three
papers~\cite{Draper:2013oza, Bagnaschi:2014rsa, Vega:2015fna}
presented updates of the traditional EFT calculation: the use of the
state-of-the-art results collected in ref.~\cite{Buttazzo:2013uya} for
the SM part (i.e., three-loop RGEs and two-loop EW-scale matching
conditions), together with the full one-loop and partial two-loop
matching conditions at the SUSY scale, allow for a full NLL and
partial NNLL resummation of the logarithmic
corrections.\footnote{Refs.~\cite{Hahn:2013ria, Draper:2013oza} also
  obtained analytic results for the coefficients of logarithmic terms
  in $\dmhq$ beyond three loops.}  Several public codes for the EFT
calculation of the Higgs mass in the MSSM with heavy SUSY have also
been released: \SHD~\cite{SusyHD}, based on ref.~\cite{Vega:2015fna};
{\tt MhEFT}~\cite{MhEFT}, based on refs.~\cite{Draper:2013oza,
  Lee:2015uza} and covering as well scenarios with a light
two-Higgs-doublet model (THDM); \hssusy~\cite{HSSUSY,Athron:2016fuq},
a module of {\tt FlexibleSUSY}~\cite{Athron:2014yba} with the same
essential features as the original \SHD; {\tt
  FlexibleEFTHiggs}~\cite{HSSUSY,Athron:2016fuq}, which combines a
full one-loop computation of $\mh$ with a LL resummation of the
logarithmic corrections; finally, an EFT approach similar to the one
of ref.~\cite{Athron:2016fuq} was recently implemented in {\tt
  SPheno/SARAH}~\cite{Staub:2017jnp}.

In MSSM scenarios with stop masses of several TeV, where the effects
of $\ovms$ can be safely neglected, the theoretical uncertainty of the
EFT prediction for the Higgs mass stems from missing terms of higher
orders in the loop expansion, both in the calculation of the matching
conditions at the SUSY scale and in the SM part of the calculation. In
refs.~\cite{Bagnaschi:2014rsa, Vega:2015fna} such uncertainty was
estimated to be at most $1$~GeV in a simplified MSSM scenario with
degenerate SUSY masses of $10$~TeV, $\tan\beta=20$ and vanishing
$X_t$\,, where $\mh \approx 123.5$~GeV. In such scenario, the
prediction for $\mh$ of the ``hybrid'' (i.e., fixed-order+partial NLL)
calculation of ref.~\cite{Hahn:2013ria} was about $3$~GeV higher, well
outside the theoretical uncertainty of the EFT result. In
refs.~\cite{Vega:2015fna, Lee:2015uza} it was suggested that most of
the discrepancy came from the determination of the coupling $g_t$ used
in the resummation procedure, for which ref.~\cite{Hahn:2013ria}
omitted one-loop EW and two-loop QCD effects, consistently with the
accuracy of the $\mh$ calculation in that paper. Those effects were
later included in \FH, which now also allows for a full NLL and
partial NNLL resummation of the logarithmic
corrections~\cite{Bahl:2016brp}. In the simplified MSSM scenario
mentioned above, the refinements in the resummation procedure of \FH\
reduce the discrepancy with the EFT prediction for $\mh$ to a few
hundred MeV.

As mentioned earlier, MSSM scenarios with stop masses below a couple
of TeV and large stop mixing -- which are definitely more interesting
from the point of view of LHC phenomenology -- suffer from even larger
spreads in the predictions of different codes for $\mh$. For example,
in a benchmark point with degenerate SUSY masses of $1$~TeV,
$\tan\beta=20$, and $X_t$ varied so as to maximize $\mh$, the EFT
calculation finds $\mhmax \approx 123$~GeV, whereas {\tt SoftSusy},
{\tt SuSpect} and {\tt SPheno} -- which implement the same corrections
to the Higgs masses, but differ in the determination of the running
couplings -- find $\mhmax \approx 124.5\!-\!126.5$~GeV, and the latest
version of \FH~\cite{fhcode} finds $\mhmax \approx 126\!-\!128$~GeV
(depending on the code's settings). However, in this case the
comparison between the EFT prediction for $\mh$ and the various
fixed-order (or hybrid) predictions is less straightforward than in
scenarios with multi-TeV stop masses, because there is no obvious
argument to favor one calculational approach over the others: the
$\ovms$ terms might or might not be negligible, and the logarithmic
corrections might or might not be important enough to mandate their
resummation. For all approaches, this unsatisfactory situation points
to two urgent needs: first, to improve the calculation of $\mh$ with
the inclusion of higher-order effects; second, to provide a better
estimate of the theoretical uncertainty, tailored to the ``difficult''
region of the parameter space with stop masses about $1\!-\!2$~TeV.

\bigskip

In this paper we take several steps towards an improved EFT
determination of the Higgs mass in the MSSM with heavy
superpartners. In particular, in section~\ref{sec:matching} we compute
the two-loop, ${\cal O}(g_t^6)$ contribution to the SUSY-scale
matching condition for the quartic Higgs coupling -- which was
previously known only in simplified scenarios~\cite{Espinosa:2000df,
  Draper:2013oza, Vega:2015fna} -- allowing for generic values of all
the relevant SUSY-breaking parameters. We also include the two-loop
contributions controlled by the bottom and tau Yukawa couplings,
addressing some subtleties related to the presence of potentially
large $\tan\beta$-enhanced corrections. Our new results bring the
matching condition for the quartic Higgs coupling to the same level,
in terms of an expansion in coupling constants, as the two-loop
Higgs-mass calculations in {\tt SoftSusy}, {\tt SuSpect} and {\tt
  SPheno}. In section~\ref{sec:dim6} we study instead the effects of a
representative subset of dimension-six operators in the EFT.  We
obtain both an improvement in our prediction for $\mh$ in scenarios
with stop masses about $1\!-\!2$~TeV and a more-realistic estimate of
the theoretical uncertainty associated to $\ovms$ effects.
The results presented in this paper have been implemented in modified
versions of the codes \SHD~\cite{SusyHD} and \hssusy~\cite{HSSUSY}.
All the analytic formulae that proved too lengthy to be printed here
are available upon request in electronic form.

\section{Two-loop matching of the quartic Higgs coupling}

\label{sec:matching}

In this section we describe our calculation of the two-loop matching
condition for the quartic Higgs coupling. We consider a setup in which
all SUSY particles as well as a linear combination of the two Higgs
doublets of the MSSM are integrated out at a common renormalization
scale $Q\approx\MS$, so that the EFT valid below the matching scale is
just the SM. Using the conventions outlined in section 2 of
ref.~\cite{Bagnaschi:2014rsa}, the two-loop matching condition for the
quartic coupling of the SM-like Higgs doublet $H$ takes the form
\beq
\label{looplam}
\lambda (Q)= \frac14\left[g^2(Q)+ g^{\prime\,2}(Q)\right] 
\cos^22\beta 
~+~ \Delta \lambda^{1\ell}
~+~ \Delta \lambda^{2\ell}~,  
\eeq
where $g$ and $g^\prime$ are the EW gauge couplings, $\beta$ can be
interpreted as the angle that rotates the two original MSSM doublets
into a light doublet $H$ and a massive doublet $A$, and
$\Delta \lambda^{n\ell}$ is the $n$-loop threshold correction to the
quartic coupling arising from integrating out the heavy particles at
the scale $\MS$.  The contributions to $\Delta \lambda^{1\ell}$
controlled by the EW gauge couplings and by the top Yukawa coupling,
for generic values of all SUSY parameters, were given in
ref.~\cite{Bagnaschi:2014rsa}, completing and correcting earlier
results of refs.~\cite {Bernal:2007uv, Giudice:2011cg}. For
completeness, we report in the appendix the full result for the
one-loop contributions of heavy scalars, including also terms
controlled by the bottom and tau Yukawa couplings.  However, the only
one-loop contributions relevant to our computation of the two-loop
threshold correction, where we will consider the ``gaugeless'' limit
$g=g^\prime=0$, are those proportional to the fourth power of a
third-family Yukawa coupling, which read:

\beq
\label{oneloop}
\Delta \lambda^{g_f^4} ~=~ \sum_{f=t,b,\tau}~
\frac{\hat g_f^4\, N^f_c}{(4\pi)^2}\,\left\{
\ln\frac{\mfl^2\mfr^2}{Q^4} 
\,+\, 2  \,\wt X_f
\left[\wt F_1 (x_f) \,-\,\frac{\wt X_f}{12}
\,\wt F_2 (x_f)\right]\right\}~,
\eeq
where by $\hat g_f$ we denote SM-like Yukawa
couplings,\footnote{Beyond tree level, we must distinguish these
  couplings from the proper Yukawa couplings of the SM, denoted as
  $g_f\,$, and specify a renormalization prescription for the angle
  $\beta$.} related to their MSSM counterparts $\hat y_f$ by
$\hat g_t=\hat y_t\,\sin\beta\,$, $\hat g_b = \hat y_b \,\cos\beta$
and $\hat g_\tau = \hat y_\tau \,\cos\beta\,$. Moreover, for each
fermion species $f$: $N^f_c$ is the number of colors; $(\mfl,\mfr)$
are the soft SUSY-breaking sfermion masses, i.e.~$(m_{Q_3},m_{U_3})$,
$(m_{Q_3},m_{D_3})$ and $(m_{L_3},m_{E_3})$ for stops, sbottoms and
staus, respectively; $\wt X_f = X_f^2/(\mfl\mfr)$, where
$X_f = A_f-\mu \,r_f$, $A_f$ is the trilinear soft SUSY-breaking
Higgs-sfermion coupling, $\mu$ is the Higgs mass term in the MSSM
superpotential, $r_t = \cot\beta$ and $r_b=r_{\tau}=\tb\,$;
$x_f = \mfl/\mfr\,$; finally, the loop functions $\wt{F}_1$ and
$\wt{F}_2$ are defined in appendix~A of ref.~\cite{Bagnaschi:2014rsa}.

For what concerns the two-loop threshold correction
$\Delta \lambda^{2\ell}$, simplified results for the
${\cal O}(g_t^4\,g_3^2)$ and ${\cal O}(g_t^6)$ contributions, valid in
the limit $m_{Q_3}=m_{U_3}=\ma=\mg\equiv M_S$ (where $\ma$ is the mass
of the heavy Higgs doublet and $\mg$ is the gluino mass), were made
available as far back as in ref.~\cite{Espinosa:2000df}. Among the
recent EFT analyses, refs.~\cite{Bagnaschi:2014rsa,Vega:2015fna}
obtained formulae for the ${\cal O}(g_t^4\,g_3^2)$ contributions valid
for arbitrary values of all the relevant SUSY-breaking parameters. The
${\cal O}(g_t^6)$ contributions, on the other hand, were neglected in
ref.~\cite{Bagnaschi:2014rsa}, while they were included in
refs.~\cite{Draper:2013oza,Vega:2015fna} only through simplified
formulae derived from those of ref.~\cite{Espinosa:2000df}. In this
paper we extend the calculations of refs.~\cite{Bagnaschi:2014rsa,
  Vega:2015fna} to obtain all contributions to
$\Delta \lambda^{2\ell}$ controlled only by the third-family Yukawa
couplings, again for arbitrary values of all the relevant
SUSY-breaking parameters. Besides improving our knowledge of the
${\cal O}(g_t^6)$ contributions from two-loop diagrams involving
stops, this allows us to properly account for sbottom and stau
contributions that can become relevant at large values of $\tb$. We
also discuss how to obtain the ${\cal O}(g_b^4\,g_3^2)$ contributions
from the known results for the ${\cal O}(g_t^4\,g_3^2)$
ones. Altogether, our results amount to a complete determination of
$\Delta \lambda^{2\ell}$ in the limit of vanishing EW gauge (and
first-two-generation Yukawa) couplings.

\subsection{Outline of the calculation}
\label{sec:lambda2loop}

The two-loop, Yukawa-induced threshold correction to the quartic Higgs
coupling $\lambda$ at the matching scale $Q$ can be expressed as
\beq
\label{effpot}
\Delta\lambda^{2\ell} ~ = ~ \frac12\,\left.
{\frac{\partial^4 \Delta V^{2\ell,\,\nonSM}}
{\partial^2H^\dagger\partial^2 H}}\,\right|_{H=0} 
\!\!+~ \Delta\lambda^{{\rm shift},\,f}
~+~ \Delta\lambda^{{\rm shift},\,\tilde f}~,
\eeq
where by $\Delta V^{2\ell,\,\nonSM}$ we denote the contribution to
the MSSM scalar potential from two-loop diagrams involving sfermions
that interact with themselves, with Higgs doublets or with matter
fermions and higgsinos only through the third-family Yukawa couplings,
as well as from two-loop diagrams involving only the heavy Higgs doublet
and matter fermions. The last two terms in eq.~(\ref{effpot}) contain
additional two-loop contributions that will be described below.
In the following we will focus on the contributions to
$\Delta\lambda^{2\ell}$ that involve the top and bottom Yukawa
couplings, and comment only briefly on the inclusion of the
contributions that involve the tau Yukawa coupling, which are in
general much smaller. However, we stress that the results that we
implemented in \SHD~\cite{SusyHD} and \hssusy~\cite{HSSUSY} (and that
we make available upon request) do include the tau-Yukawa
contributions through two loops.

In the gaugeless limit adopted in our calculation, the
field-dependent mass spectrum of the particles that enter the relevant
two-loop diagrams simplifies considerably: we can approximate the
masses of the lightest Higgs scalar and of the would-be Goldstone
bosons to zero, and the masses of all components (scalar, pseudoscalar
and charged) of the heavy Higgs doublet to $\ma^2$; the charged and
neutral components of the two higgsino doublets combine into Dirac
spinors with degenerate mass eigenvalues $|\mu|^2$; the tree-level
mixing angle in the CP-even sector is just $\alpha = \beta - \pi/2$.
For the contributions to $\Delta V^{2\ell,\,\nonSM}$ that involve
the top and bottom Yukawa couplings, we adapt the results used for the
effective-potential calculation of the MSSM Higgs masses in
ref.~\cite{Dedes:2003km}.\footnote{We compared our result for the top
  and bottom Yukawa contribution to $\Delta V^{2\ell,\,\nonSM}$ with
  the one obtained by imposing the gaugeless limit and removing the
  SM-like contribution in eq.~(D.6) of ref.~\cite{Espinosa:2000df}. We
  find agreement except for the overall sign of the next-to-last line
  of that equation.} To compute the fourth derivative of the effective
potential entering eq.~(\ref{effpot}) we follow the approach outlined
in section 2.3 of ref.~\cite{Bagnaschi:2014rsa}: we express the stop
and sbottom masses and mixing angles as functions of field-dependent
top and bottom masses, $m_t = \hat g_t\, |H|$ and
$m_b = \hat g_b \,|H|$, and obtain
\bea
\label{derivs}
\left.
\frac{\partial^4 \Delta V^{2\ell,\,\nonSM}}
{\partial^2H^\dagger\partial^2 H}\,\right|_{H=0}  &=&
\biggr[~ 
\hat g_t^4\,\left( 2 \,V_{tt}^{(2)} \,+\, 4\,m_t^2 \,V_{ttt}^{(3)}
\,+\, m_t^4 \,V_{tttt}^{(4)}\right)\nn\\
&& ~~+ \hat g_t^2\,\hat g_b^2
\left( 2 \,V_{tb}^{(2)} \,+\, 12\,m_t^2 \,V_{ttb}^{(3)}
\,+\, 4\,m_t^4 \,V_{tttb}^{(4)}\,+\, 3\,m_t^2 \,m_b^2\, V_{ttbb}^{(4)}
\right)\biggr]_{m_t,m_b\,\rightarrow \,0}\nn\\
&+& \biggr[\,t \,\longleftrightarrow\, b\biggr]~,
\eea
where the term in the last line is obtained from the terms in the
first two lines by swapping top and bottom, and we used the shortcuts
\beq
\label{shortcut}
V^{(k)}_{q_1\dots\, q_k} ~=~ 
\frac {d^k \Delta V^{2\ell,\,\nonSM}}{d m^2_{q_1}\dots\,d m^2_{q_k}}~.
\eeq
The derivatives of the field-dependent stop and sbottom parameters and
the limit of vanishing top and bottom masses in eq.~(\ref{derivs}) are
obtained as described in ref.~\cite{Bagnaschi:2014rsa}. As in the case
of the ${\cal O}(g_t^4\,g_3^2)$ contributions, we find that the fourth
derivative of the two-loop effective potential contains terms
proportional to $\ln(m_q^2/Q^2)$, which would diverge for vanishing
quark masses but cancel out against similar terms in the contribution
denoted as $\Delta\lambda^{{\rm shift},\,f}$ in
eq.~(\ref{effpot}). Indeed, above the matching scale the one-loop
contribution to the quartic Higgs coupling from box diagrams with a
top or bottom quark,
\beq
\label{deltaq}
\delta \lambda^{ g_q^4,\, q} ~=~ - \sum_{q=t,b} 
\frac{\hat g_q^4\,N_c}{(4\pi)^2}\,
\left(2\,\ln\frac{m_q^2}{Q^2} + 3 \right)~,
\eeq
is expressed in terms of the MSSM couplings $\hat g_q$, whereas below
the matching scale the same contribution is expressed in terms of the
SM couplings $g_q$. To properly compute the two-loop, Yukawa-only part
of the matching condition for the quartic Higgs coupling, we must
re-express the MSSM couplings entering $\delta \lambda^{ g_q^4,\, q}$
above the matching scale (including those implicit in $m_q$) according
to
$\hat g_q \,\rightarrow g_q\,(1+\Delta g_q^{\scriptscriptstyle Y})$,
where $\Delta g_q^{\scriptscriptstyle Y}$ denotes the terms controlled
by the Yukawa couplings in the threshold correction to $g_q$. In
particular, we find for the top and bottom Yukawa couplings
\bea
\label{deltagt}
\Delta g_t^{\scriptscriptstyle Y} &=& 
- \frac{\hat g_t^2}{(4\pi)^2\sin^2\beta} \,\left[ \,
\frac{3}{4}\,\ln \frac{\mu^2}{Q^2}
\,+\,\frac{3}{8}\,\cos^2\beta\, \left( 2\, \ln \frac{\ma^2}{Q^2} -1\right)  
\,+\,\wt{F}_6 \left( \frac{m_{Q_3}}{\mu}\right) 
\,+\, \frac{1}{2}\, \wt{F}_6 \left(
  \frac{m_{U_3}}{\mu} \right)\, \right]\nn\\[2mm]
&&- \frac{\hat g_b^2}{(4\pi)^2\cos^2\beta} \,\left[ \,
\frac{1}{4}\,\ln \frac{\mu^2}{Q^2}
\,+\,\frac{1}{8}\,\sin^2\beta\, \left( 2\, \ln \frac{\ma^2}{Q^2} -1\right)
\,+\,\cos^2\beta\, \left(\ln \frac{\ma^2}{Q^2} -1\right)
\right.\nn\\[1mm]
&&~~~~~~~~~~~~~~~~~~~~\left.\,+\, \frac{1}{2}\, \wt{F}_6 \left(
  \frac{m_{D_3}}{\mu} \right) \,+\,
\frac{X_b\,\cot\beta}{2\,\mu}\,\wt{F}_9 \left(
   \frac{m_{Q_3}}{\mu},\frac{m_{D_3}}{\mu} \right)\right]
~-~ \frac{\delta Z_\smallH^{\tilde q}}{2}~,\\[4mm]
\label{deltagb}
\Delta g_b^{\scriptscriptstyle Y} &=& 
- \frac{\hat g_b^2}{(4\pi)^2\cos^2\beta} \,\left[ \,
\frac{3}{4}\,\ln \frac{\mu^2}{Q^2}
\,+\,\frac{3}{8}\,\sin^2\beta\, \left( 2\, \ln \frac{\ma^2}{Q^2} -1\right)  
\,+\,\wt{F}_6 \left( \frac{m_{Q_3}}{\mu}\right) 
\,+\, \frac{1}{2}\, \wt{F}_6 \left(
  \frac{m_{D_3}}{\mu} \right)\, \right]\nn\\[2mm]
&&- \frac{\hat g_t^2}{(4\pi)^2\sin^2\beta} \,\left[ \,
\frac{1}{4}\,\ln \frac{\mu^2}{Q^2}
\,+\,\frac{1}{8}\,\cos^2\beta\, \left( 2\, \ln \frac{\ma^2}{Q^2} -1\right)
\,+\,\sin^2\beta\, \left(\ln \frac{\ma^2}{Q^2} -1\right)
\right.\nn\\[1mm]
&&~~~~~~~~~~~~~~~~~~~~\left.\,+\, \frac{1}{2}\, \wt{F}_6 \left(
  \frac{m_{U_3}}{\mu} \right) \,+\,
\frac{X_t\,\tan\beta}{2\,\mu}\,\wt{F}_9 \left(
   \frac{m_{Q_3}}{\mu},\frac{m_{U_3}}{\mu} \right)\right]
~-~ \frac{\delta Z_\smallH^{\tilde q}}{2}~,
\eea
where the last term on the right-hand side of each equation reads, in
a notation analogous to the one of eq.~(\ref{oneloop}),
\beq
\label{WFR}
\delta Z_\smallH^{\tilde q} ~=~ 
-\sum_{q=t,b} \frac{\hat g_q^2\, N_c}{(4\pi)^2}~
\frac{\wt X_q}{6}~ \wt{F}_5 (x_q)~,
\eeq
and corresponds to the threshold correction to the light-Higgs WFR
arising from squark loops.
The loop functions $\wt{F}_5$, $\wt{F}_6$ and $\wt{F}_9$ are defined
in appendix~A of ref.~\cite{Bagnaschi:2014rsa}. We also remark that
eqs.~(\ref{deltagt})--(\ref{WFR}) assume that the angle $\beta$
entering the relations between the SM-like couplings $\hat g_q$ and
their MSSM counterparts $\hat y_q$ is renormalized as described in
section 2.2 of ref.~\cite{Bagnaschi:2014rsa}, removing entirely the
contributions of the off-diagonal WFR of the Higgs doublets.
Combining the effects of the shifts in the Yukawa couplings with the
renormalization of the Higgs fields (keeping into account also the
field-dependent quark masses in the logarithms) we obtain the total
contribution to $\Delta\lambda^{2\ell}$ arising from the quark-box
diagrams of eq.~(\ref{deltaq}),
\beq
\label{dshiftq}
\Delta\lambda^{{\rm shift},\,f} ~=~
- \sum_{q=t,b} 
\frac{\hat g_q^4\,N_c}{(4\pi)^2}\,
\left(2\,\ln\frac{m_q^2}{Q^2} + 4 \right)
\left(4\,\Delta g_q^{\scriptscriptstyle Y} 
+ 2\,\delta Z_\smallH^{\tilde q}\right)~,
\eeq
which cancels the logarithmic dependence on the quark masses of the
derivatives of $\Delta V^{2\ell,\,\nonSM}$. We checked that the
contributions in eq.~(\ref{derivs}) that involve more than two
derivatives of the two-loop effective potential cancel out completely
against the shift of the corresponding contributions in the one-loop
part -- namely, the non-logarithmic term in the right-hand side of
eq.~(\ref{deltaq}) -- so that the final result for
$\Delta\lambda^{2\ell}$ can be related to the two-loop correction to
the light-Higgs mass. This is the same ``decoupling'' property found
in ref.~\cite{Bagnaschi:2014rsa} for the ${\cal O}(g_t^4\,g_3^2)$ part
of $\Delta\lambda^{2\ell}$.  Finally, it can be inferred from
eqs.~(\ref{deltagt})--(\ref{dshiftq}) that the contribution of
$\delta Z_\smallH^{\tilde q}$ cancels out of
$\Delta\lambda^{{\rm shift},\,f}$.

The last term in eq.~(\ref{effpot}),
$\Delta\lambda^{{\rm shift},\,\tilde f}$, arises from shifts in the
sfermion contribution to the one-loop matching condition for the
quartic Higgs coupling, eq.~(\ref{oneloop}). In particular, it
contains terms arising from the WFR of the Higgs fields, which are not
captured by the derivatives of $\Delta V^{2\ell,\,\nonSM}$, plus
additional contributions that arise if we express the one-loop
threshold correction in eq.~(\ref{oneloop}) in terms of the SM Yukawa
couplings, $g_q$, instead of the MSSM ones, $\hat g_q$. We remark here
that, while the shift of the Yukawa couplings in the quark-box
diagrams of eq.~(\ref{deltaq}) is required for a consistent two-loop
matching of the quartic Higgs coupling, an analogous shift in the
squark contribution of eq.~(\ref{oneloop}) is to some extent a matter
of choice. In refs.~\cite{Bagnaschi:2014rsa, Vega:2015fna} the top
Yukawa coupling entering the one-loop part of the threshold correction
to the quartic Higgs coupling was interpreted as the SM one. Applying
that choice to both the top and bottom Yukawa couplings, we would find
\beq
\label{dshiftsq}
\Delta\lambda^{{\rm shift},\,\tilde f} ~=~
\sum_{q=t,b}~
\frac{\hat g_q^4\, N_c}{(4\pi)^2}\,\left\{
\ln\frac{\mql^2\mqr^2}{Q^4} 
\,+\, 2 \, \wt X_q
\left[\wt F_1 (x_q) \,-\,\frac{\wt X_q}{12}
\,\wt F_2 (x_q)\right]\right\}
\left(4\,\Delta g_q^{\scriptscriptstyle Y} 
+ 2\,\delta Z_\smallH^{\tilde q}\right)
~,
\eeq
where again the contributions of the WFR of the Higgs fields cancel
out against analogous terms in the shifts of the Yukawa
couplings. After including in $\Delta\lambda^{2\ell}$ the shifts in
eqs.~(\ref{dshiftq}) and (\ref{dshiftsq}), we checked that, in the
limit of $g_b=0$ and $m_{Q_3}=m_{U_3}=\ma\equiv M_S$, the
${\cal O}(g_t^6)$ part of $\Delta\lambda^{2\ell}$ coincides with the
simplified result given in eq.~(21) of ref.~\cite{Vega:2015fna}.

\bigskip

On the other hand, it is well known~\cite{Hempfling:1993kv,
  Hall:1993gn, Carena:1994bv} that the relation between the bottom
Yukawa coupling of the SM and its MSSM counterpart is subject to
potentially large corrections enhanced by $\tb$, which, in the
gaugeless limit, arise from diagrams involving either gluino-sbottom
or higgsino-stop loops. As discussed, e.g., in
ref.~\cite{Carena:1999py}, these $\tb$-enhanced terms can be
``resummed'' in the $\drbar$-renormalized coupling of the MSSM by
expressing it as
\beq
\label{resum}
\hat g_b(Q) ~=~ \frac{ g_b(Q)}{1- \left(\Delta g_b^{s} 
+ \Delta g_b^{\scriptscriptstyle Y}\right)}~,
\eeq
where $g_b(Q)$ is the $\msbar$-renormalized coupling of the SM,
extracted at the EW scale from the bottom mass and evolved up to the
matching scale $Q$ with SM RGEs, $\Delta g_b^{\scriptscriptstyle Y}$ is
given in eq.~(\ref{deltagb}), and
\beq
\label{deltagb_s}
\Delta g_b^{s} ~=~ - \frac{g_3^2\,C_F}{(4\pi)^2}\,\left[
1 \,+\,\ln\frac{\mg^2}{Q^2} 
\,+\, \wt F_6\left(\frac{m_{Q_3}}{\mg}\right)
\,+\, \wt F_6\left(\frac{m_{D_3}}{\mg}\right)
\,-\, \frac{X_b}{\mg}\,
\wt F_9\left(\frac{m_{Q_3}}{\mg},\frac{m_{D_3}}{\mg}\right)\right]~,
\eeq
where $C_F=4/3$ is a color factor, and we recall that
$X_b=A_b-\mu\,\tb$. 
In contrast with our treatment of the top Yukawa coupling, we will
therefore choose to interpret the bottom Yukawa coupling entering the
one-loop part of the threshold correction to the quartic Higgs
coupling as the MSSM one, in order to absorb the $\tb$-enhanced
effects directly in $\Delta\lambda^{1\ell}$. We recall that a similar
approach was discussed in refs.~\cite{Brignole:2002bz, Dedes:2003km,
  Heinemeyer:2004xw} in the context of the fixed-order calculation of
the Higgs masses in the MSSM.

With our choice for the bottom Yukawa coupling entering
$\Delta\lambda^{1\ell}$, we must omit the term
$4\,\Delta g_b^{\scriptscriptstyle Y}$ in the formula for
$\Delta\lambda^{{\rm shift},\,\tilde f}$, see eq.~(\ref{dshiftsq}),
when computing the contributions to $\Delta\lambda^{2\ell}$ controlled
only by the top and bottom Yukawa couplings. Concerning the
${\cal O}(g_b^4\,g_3^2)$ contributions, they can be obtained from the
${\cal O}(g_t^4\,g_3^2)$ contributions computed in
refs.~\cite{Bagnaschi:2014rsa, Vega:2015fna} via
\bea
\label{2loopstr}
\Delta\lambda^{g_b^4\,g_3^2} &=& \Delta\lambda^{g_t^4\,g_3^2} 
~[\,t \,\rightarrow\, b\,]\nn\\[1mm]
&-& 
4\,\Delta g_b^s~\frac{\hat g_b^4\, N_c}{(4\pi)^2}\,\left\{
\ln\frac{m_{Q_3}^2m_{D_3}^2}{Q^4} 
\,+\, 2 \, \wt X_b
\left[\wt F_1 (x_b) \,-\,\frac{\wt X_b}{12}
\,\wt F_2 (x_b)\right]\right\}~,
\eea
where the notation $[\,t \,\rightarrow\, b\,]$ in the first line
represents the replacements $g_t \rightarrow g_b$,
$X_t \rightarrow X_b$ and $m_{U_3} \rightarrow m_{D_3}$ in the
formulae for the ${\cal O}(g_t^4\,g_3^2)$ contributions. We note that,
in practice, our choice removes from $\Delta\lambda^{2\ell}$
potentially large terms characterized by a higher power of $\tb$ than
of $\hat g_b$, i.e.~terms scaling like
$\hat g_b^4\,g_3^2\,\tan^5\!\beta$ or like
$\hat g_b^4\,\hat g_t^2\,\tan^5\!\beta$.
 
\bigskip

We now comment on the inclusion of the contributions to
$\Delta\lambda^{2\ell}$ controlled by the tau Yukawa coupling. The
two-loop contributions of ${\cal O}(g_\tau^6)$, i.e.~those involving
only the tau Yukawa coupling, do not require a separate calculation,
since they can be obtained from the top-only, ${\cal O}(g_t^6)$ ones
via the replacements $g_t \rightarrow g_\tau$,
$A_t\rightarrow A_{\tau}$, $N_c\rightarrow 1$,
$m_{Q_3} \rightarrow m_{L_3}$, $m_{U_3} \rightarrow m_{E_3}$ and
$\cos\beta \leftrightarrow \sin\beta$ (see also
ref.~\cite{Dedes:2003km}). Indeed, as long as we neglect the EW gauge
couplings, the threshold correction to the tau Yukawa coupling does
not contain any $\tb$-enhanced terms, and reads
\beq
\label{deltagtau}
\Delta g_\tau ~=~
- \frac{\hat g_\tau^2}{(4\pi)^2\cos^2\beta} \,\left[ \,
\frac{3}{4}\,\ln \frac{\mu^2}{Q^2}
\,+\,\frac{3}{8}\,\sin^2\beta\, \left( 2\, \ln \frac{\ma^2}{Q^2} -1\right)  
+\,\wt{F}_6 \left( \frac{m_{L_3}}{\mu}\right) 
+\, \frac{1}{2}\, \wt{F}_6 \left(
  \frac{m_{E_3}}{\mu} \right) \right]
-~ \frac{\delta Z_\smallH^{\tilde f}}{2}~,
\eeq
where the sfermion contribution to the Higgs WFR,
$\delta Z_\smallH^{\tilde f}$, is obtained including also the stau
contribution (with $N_c=1$) in eq.~(\ref{WFR}). We can therefore treat
the tau Yukawa coupling in the same way as the top one, expressing the
stau contribution to $\Delta\lambda^{1\ell}$ in terms of the SM
coupling $g_\tau$.
In addition, ``mixed'' contributions to the two-loop effective
potential controlled by both the tau and bottom Yukawa couplings arise
from diagrams that involve the quartic sbottom-stau coupling, see
appendix B of ref.~\cite{Allanach:2004rh}. The corresponding
contributions to $\Delta\lambda^{2\ell}$ can be obtained directly from
the derivatives of the effective potential (without additional shifts)
with the procedure outlined around eq.~(\ref{derivs}), after replacing
$t \rightarrow \tau$ in the latter.
Finally, the choice of using the MSSM coupling $\hat g_b$ in
$\Delta\lambda^{1\ell}$ spoils the cancellation of Higgs WFR effects
in $\Delta\lambda^{{\rm shift},\,\tilde f}$, see
eq.~(\ref{dshiftsq}). As a result, when we take into account the
${\cal O}(g_\tau^2)$ contribution from stau loops in
$\delta Z_\smallH^{\tilde f}$, we find an additional
${\cal O}(g_b^4\,g_\tau^2)$ contribution to $\Delta\lambda^{2\ell}$.

\subsection{Numerical examples}
\label{sec:tlnumbers}

We now provide some illustration of the numerical impact of the newly
computed two-loop corrections to the quartic Higgs coupling. To this
purpose, we implemented those corrections in modified versions of the
codes \SHD~\cite{SusyHD} and \hssusy~\cite{HSSUSY}. All plots
presented in this section were produced with \hssusy, but we checked
that fully analogous plots can be obtained with \SHD. Small
discrepancies in the predictions for $\mh$ arise from differences in
the calculations implemented in the two codes, as discussed in section
2.3 of ref.~\cite{Athron:2016fuq}, but they do not affect the
qualitative behavior and relative importance of the new two-loop
corrections.
The SM input parameters used for \hssusy\ in our studies, which we
take from ref.~\cite{Olive:2016xmw}, are: the Fermi constant
$G_F=1.16638\!\times\!10^{-5}$~GeV$^{-2}$; the Z boson mass
$\MZ=91.1876$~GeV; the pole top mass $M_t^{\rm pole} = 173.21$~GeV;
the $\msbar$-renormalized bottom mass $\mb(\mb)=4.18$~GeV; the tau
mass $m_\tau = 1777$~MeV; finally, the strong and electromagnetic
coupling constants in the five-flavor $\msbar$ scheme,
$\alpha_s(\MZ)=0.1181$ and $\alpha(\MZ)=1/127.950$.

\bigskip

To start with, we omit all contributions to $\Delta\lambda^{2\ell}$
controlled by the bottom and tau Yukawa couplings, and focus on the
effect of extending the contributions controlled by the top Yukawa
coupling to generic values of the relevant SUSY-breaking parameters.
We consider a scenario in which all SUSY-particle masses are larger
than one TeV, but the stop masses are not degenerate. In particular,
we take $m_{U_3}= 1.5$~TeV and $m_{Q_3}= \kappa\,m_{U_3}$, where
$\kappa$ is a scaling parameter that we vary in the range
$1\leq\kappa\leq4$. We also take $\mg=\ma = m_{U_3}$,
$\,\mu = 4\,m_{U_3}$ and $\tan\beta=20$, and we fix $A_t$ via the
``maximal'' stop mixing condition
$A_t-\mu\,\cot\beta = (6 \,m_{Q_3}\,m_{U_3})^{1/2}$. For the remaining
MSSM parameters, which affect the one-loop part of the calculation, we
set all sfermion masses other than those of the stops, as well as the
EW gaugino masses, equal to $m_{U_3}$, and we take
$A_b=A_{\tau}=A_t$. All of the MSSM parameters listed above -- with
the exception of $\tb$, which is defined as described in section 2.2
of ref.~\cite{Bagnaschi:2014rsa} -- are interpreted as
$\drbar$-renormalized parameters at the scale
$Q=(m_{Q_3}\,m_{U_3})^{1/2}$.

\begin{figure}[t]
\begin{center}
\includegraphics[width=0.45\textwidth]{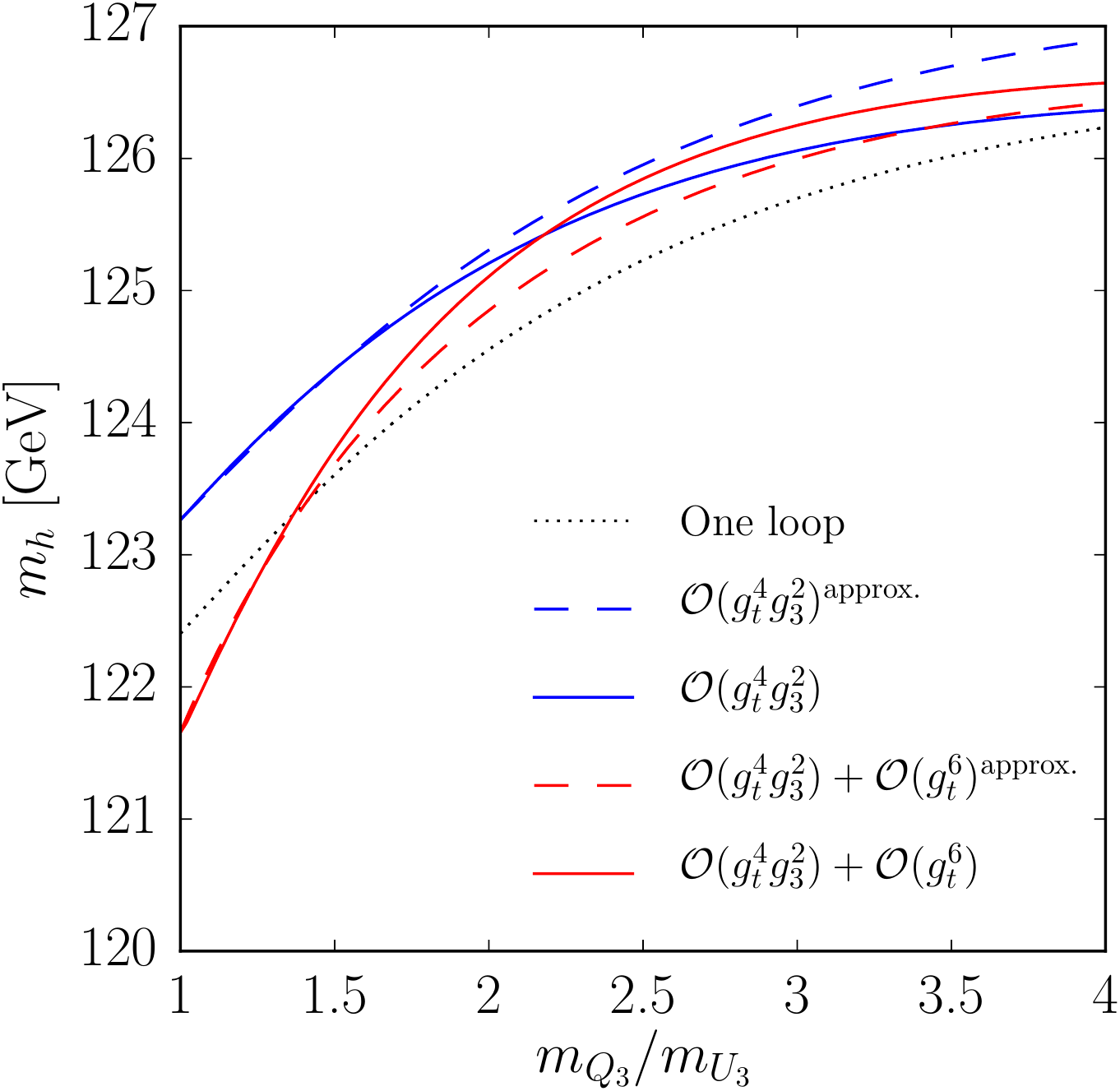}~~~~~~
\includegraphics[width=0.45\textwidth]{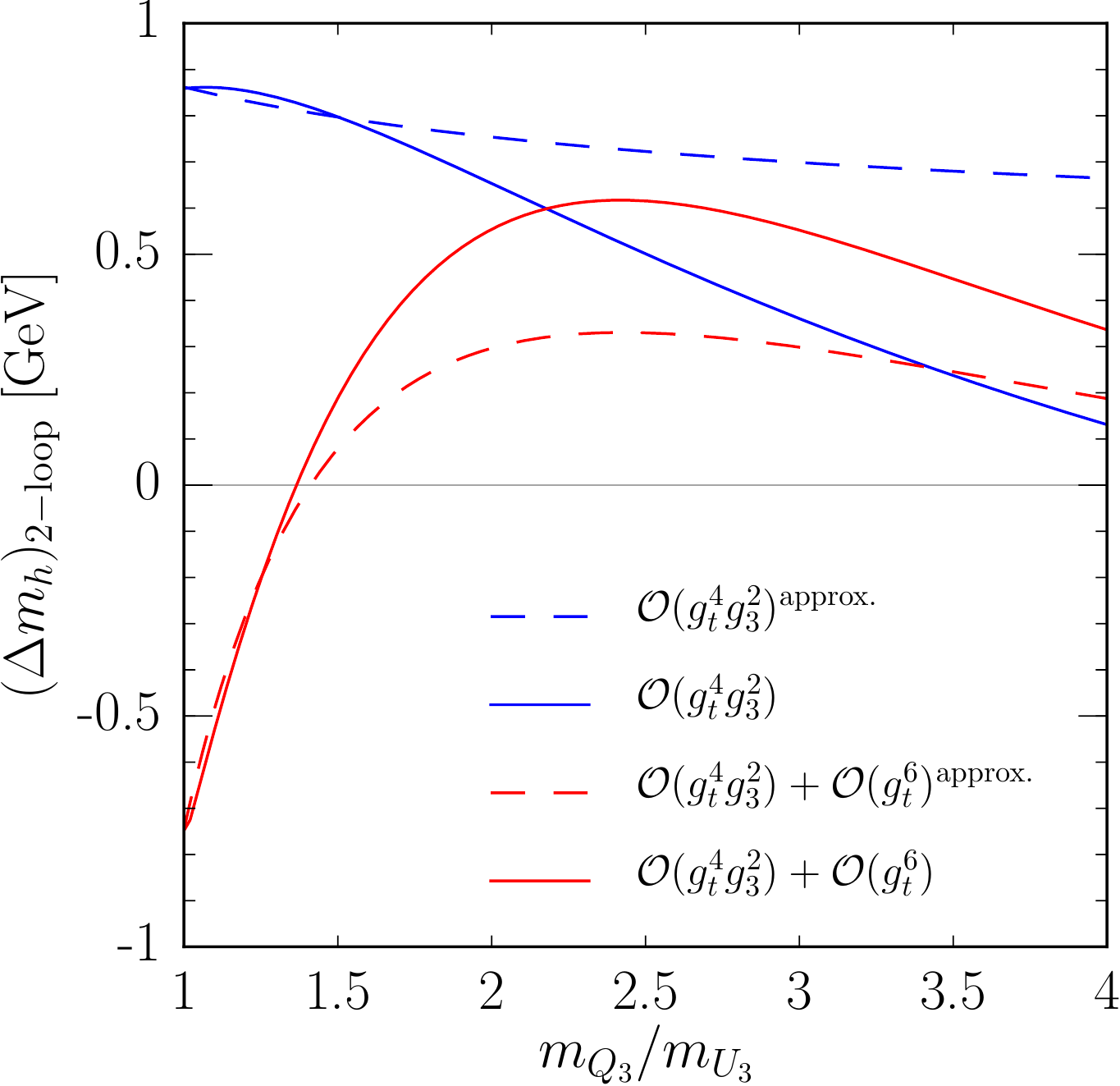}
\vspace*{2mm}
  \caption{\em Effects of the top-Yukawa contributions to
    $\Delta\lambda^{2\ell}$ in a scenario with non-degenerate SUSY
    masses, compared with approximate results obtained with degenerate
    masses. The left plot shows the predictions for $\mh$ as a
    function of the ratio of soft SUSY-breaking stop masses
    $m_{Q_3}/m_{U_3}\,$, while the right plot shows the shifts in
    $\mh$ induced by the two-loop contributions alone. The choices of
    MSSM parameters and the meaning of the different curves are
    described in the text.}
  \label{fig:2ltop}
\end{center}
\end{figure}

In figure~\ref{fig:2ltop} we compare the predictions for $\mh$
obtained with the ``exact'' (i.e., valid for generic SUSY masses)
formulae for the top-Yukawa contributions to $\Delta\lambda^{2\ell}$
with ``approximate'' predictions obtained by replacing the scalar and
gluino masses of our scenario with the degenerate masses
$m_{Q_3}^\prime = m_{U_3}^\prime = \ma^\prime = \mg^\prime =
(m_{Q_3}\,m_{U_3})^{1/2}$, and then using for the
${\cal O}(g_t^4\,g_3^2)$ and ${\cal O}(g_t^6)$ contributions to
$\Delta\lambda^{2\ell}$ the simplified formulae given in
refs.~\cite{Bagnaschi:2014rsa} and~\cite{Vega:2015fna}, respectively.
In particular, the dotted black line in the left plot of
figure~\ref{fig:2ltop} represents the prediction for $\mh$, as a
function of the stop mass ratio $\kappa = m_{Q_3}/m_{U_3}\,$, obtained
by neglecting all two-loop contributions to the matching of the
quartic Higgs coupling, and using the exact results from
refs.~\cite{Bagnaschi:2014rsa, Vega:2015fna} for the one-loop
contributions; the dashed blue line includes also the simplified
${\cal O}(g_t^4\,g_3^2)$ contributions given in eq.~(36) of
ref.~\cite{Bagnaschi:2014rsa}; the solid blue line includes instead
the exact ${\cal O}(g_t^4\,g_3^2)$ contributions from
refs.~\cite{Bagnaschi:2014rsa, Vega:2015fna}; the dashed red line
includes, on top of the exact ${\cal O}(g_t^4\,g_3^2)$ contributions,
the simplified ${\cal O}(g_t^6)$ contributions given in eq.~(21) of
ref.~\cite{Vega:2015fna}; finally, the solid red line includes instead
the exact ${\cal O}(g_t^6)$ contributions derived in this paper. In
the right plot of of figure~\ref{fig:2ltop} we show for clarity the
effect on $\mh$ of the different implementations of the two-loop
corrections alone, i.e.~we show the difference between the (dashed or
solid, blue or red) two-loop lines and the (dotted, black) one-loop
line of the left plot. The meaning of each line in the right plot
mirrors the one of the corresponding line in the left plot.

Figure~\ref{fig:2ltop} confirms that, as already noticed in
refs.~\cite{Bagnaschi:2014rsa, Vega:2015fna}, the overall effect of
the top-Yukawa contributions to $\Delta\lambda^{2\ell}$ on the EFT
predictions for $\mh$ in scenarios with multi-TeV stop masses is
rather small, typically less than one GeV. However, the comparison
between the dashed and solid lines in the plots of
figure~\ref{fig:2ltop} shows that, in scenarios with non-degenerate
mass spectra, the use of simplified formulae with an ``average'' SUSY
mass can lead to a rather poor approximation of the exact results. In
particular, the comparison between dashed and solid blue lines shows
that by using eq.~(36) of ref.~\cite{Bagnaschi:2014rsa} for the
${\cal O}(g_t^4\,g_3^2)$ corrections we would significantly
overestimate their effect on $\mh$ when $\kappa\gtrsim 2$ in our
scenario. In turn, the dashed and solid red lines show that, by using
eq.~(21) of ref.~\cite{Vega:2015fna} for the ${\cal O}(g_t^6)$
corrections, we could entirely mischaracterize their effect on the
Higgs mass: between the point where the solid blue line crosses the
solid red one and the point where it crosses the dashed red one, the
approximate calculation of the ${\cal O}(g_t^6)$ corrections gives a
negative shift in $\mh$, while the exact calculation gives a positive
shift. We remark, however, that the latter finding depends on the
somewhat large value of $\mu$ adopted in our scenario: for smaller
$\mu$ the quality of the approximation for the ${\cal O}(g_t^6)$
corrections would improve.

\bigskip

\begin{figure}[t]
\begin{center}
  \includegraphics[width=0.6 \textwidth]{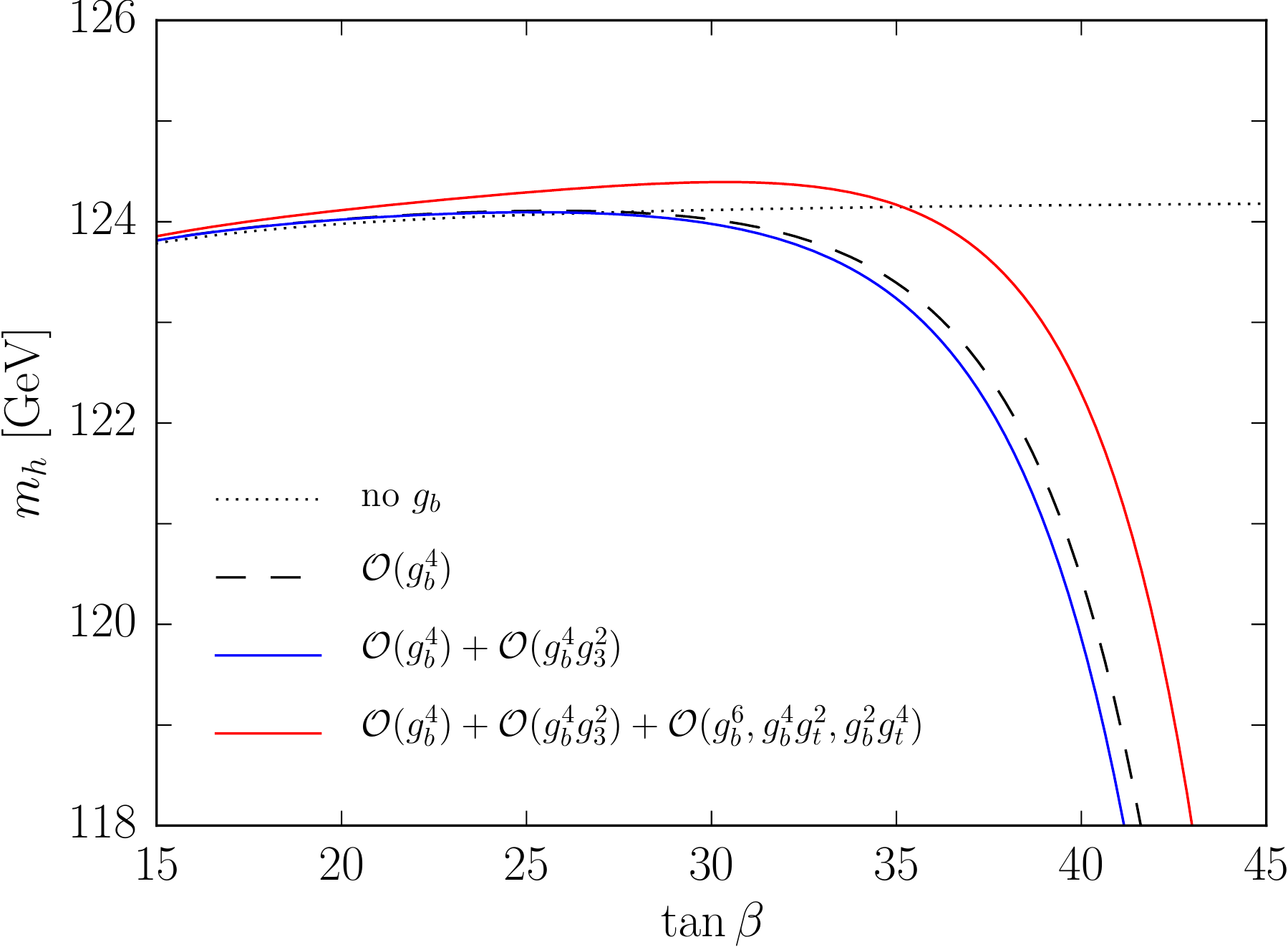}
  \caption{\em Predictions for $\mh$ as a function of $\tb$ for
    different implementations of the corrections controlled by the
    bottom Yukawa coupling. We consider an MSSM scenario with all SUSY
    masses equal to $\MS=1.5~{\rm TeV}$ except $\mg = 2.5~{\rm TeV}$, and
    with $\mu=-1.5~{\rm TeV}$, $X_t= \sqrt{6}\,\MS$ and
    $A_b=A_{\tau}=A_t$. The meaning of the different curves is
    explained in the text.}
  \label{fig:2lbot}
\vspace*{-1mm}
\end{center}
\end{figure}

We now turn our attention to the effect of the threshold corrections
to the quartic Higgs coupling controlled by the bottom Yukawa
coupling. In figure~\ref{fig:2lbot} we show the EFT prediction for
$\mh$ as a function of $\tb$, in a simplified MSSM scenario with all
soft SUSY-breaking masses of sfermions and EW gauginos, as well as the
heavy Higgs-doublet mass $\ma$, set equal to $\MS = 1.5$~TeV, while
the gluino mass is set to $\mg = 2.5$~TeV; the trilinear Higgs-stop
coupling $A_t$ is fixed by the maximal mixing condition
$A_t-\mu\,\cot\beta = \sqrt{6}\,\MS\,$, and $A_b=A_{\tau}=A_t$;
finally, we take $\mu = -1.5$~TeV, to enhance the effect of the
corrections controlled by the bottom Yukawa coupling. Indeed, negative
values of the products $\mu\,\mg$ and $\mu A_t$ ensure that $\hat g_b$
-- which we extract at the matching scale from the SM coupling $g_b$
via eq.~(\ref{resum}) -- becomes larger for increasing $\tb$, and
possibly hits a pole as the denominator on the right-hand side of
eq.~(\ref{resum}) approaches zero.  Again, all soft SUSY-breaking
parameters as well as $\mu$ are renormalized in the $\drbar$ scheme at
the matching scale $Q=\MS$.

The dotted black line in figure~\ref{fig:2lbot}, which shows very
little dependence on $\tb$, represents the prediction for $\mh$
obtained by omitting the one- and two-loop corrections to the quartic
Higgs coupling controlled by the bottom Yukawa coupling altogether;
the dashed black line includes the one-loop ${\cal O}(g_b^4)$
contribution to $\Delta\lambda^{1\ell}$, which, as discussed in
section~\ref{sec:lambda2loop}, we express in terms of the MSSM
coupling $\hat g_b\,$; the solid blue line includes also the two-loop
${\cal O}(g_b^4\,g_3^2)$ contributions to $\Delta\lambda^{2\ell}$;
finally, the solid red line includes also the two-loop
${\cal O}(g_b^6,\,g_b^4\,g_t^2,\,g_b^2\,g_t^4)$ contributions to
$\Delta\lambda^{2\ell}$.
The comparison between the dashed black line and the solid blue and
red lines shows that, when expressed in terms of the MSSM coupling
$\hat g_b$, the ${\cal O}(g_b^4)$ contribution to
$\Delta\lambda^{1\ell}$ already determines the bulk of the dependence
of $\mh$ on $\tb$. Indeed, only at rather large $\tb$, where the
dependence becomes steep, can the ${\cal O}(g_b^4\,g_3^2)$ and
${\cal O}(g_b^6,\,g_b^4\,g_t^2,\,g_b^2\,g_t^4)$ contributions to
$\Delta\lambda^{2\ell}$ shift the prediction for $\mh$ by more than
one GeV. Moreover, those corrections partially cancel out for our
choice of MSSM parameters.

Finally, we recall that the strong dependence of $\mh$ on $\tb$
depicted in figure~\ref{fig:2lbot} follows from our choice of signs
for the products $\mu\,\mg$ and $\mu A_t\,$. If both of those products
were positive instead of negative, the threshold correction
$\left(\Delta g_b^{s} + \Delta g_b^{\scriptscriptstyle Y}\right)$ in
eq.~(\ref{resum}) would suppress the MSSM coupling $\hat g_b$ -- as
well as the corresponding contributions to the quartic Higgs coupling
and, in turn, to $\mh$ -- at large values of $\tb$. If the two
products had opposite signs, the dependence of $\mh$ on $\tb$ would
hinge on whether it is $\Delta g_b^{s}$ or
$\Delta g_b^{\scriptscriptstyle Y}$ that prevails in
eq.~(\ref{resum}).


\section{On the effects of dimension-six operators}
\label{sec:dim6}

In MSSM scenarios with SUSY masses up to a couple of TeV, the effects
suppressed by powers of $v^2/\MS^2$ -- which are not accounted for
when the EFT valid below the SUSY scale involves only renormalizable
operators -- might still be relevant. In the code
\SHD~\cite{Vega:2015fna,SusyHD} the uncertainty of the prediction for
the Higgs mass associated to the omission of those effects is obtained
by multiplying the contribution to $\Delta\lambda^{1\ell}$ from each
SUSY particle by a factor\,\footnote{Note that in this paper we
  normalize the Higgs vev as $ v = \langle H^0\rangle$, with
  $v\approx174$~GeV.}  $(1\pm2\, v^2/M_i^2)\,$, where $M_i$ is that
particle's mass. In a simplified scenario with $\tb=20$, degenerate
SUSY masses $M_i\equiv\MS$ and ``maximal'' $X_t = \sqrt{6}\,\MS$, the
uncertainty arising from missing $\ovms$ effects was thus estimated in
ref.~\cite{Vega:2015fna} to be about $0.6$~GeV for $\MS=1$~TeV, and to
decrease rapidly for larger $\MS$. The total theoretical uncertainty
of the EFT prediction for $\mh$, including also the effects of missing
higher-order terms in the matching at the SUSY scale and in the SM
part of the calculation, was estimated in ref.~\cite{Vega:2015fna} to
be less than $2$~GeV for $\MS=1$~TeV, where \SHD\ finds
$\mh\approx 123$~GeV. As mentioned in section~\ref{sec:intro}, in that
scenario the predictions for the Higgs mass of various fixed-order (or
hybrid) codes differ form each other by several GeV, and in general
lie outside the estimated uncertainty of the EFT result.
In this section we aim to improve the EFT calculation of the Higgs
mass at moderate values of $\MS$ by including some of the most
important $\ovms$ effects, and to appraise the existing estimate of
the uncertainty associated to the missing ones.

\subsection{Outline of the calculation}

In the EFT framework, the effects of $\ovms$ in the predictions for
physical observables such as the Higgs mass arise from
non-renormalizable, dimension-six effective operators. The most
general dimension-six Lagrangian respecting the field content and
symmetries of the SM contains a large number of operators, see
refs.~\cite{Willenbrock:2014bja, Masso:2014xra, Pomarol:2014dya,
  Falkowski:2015fla, David:2015waa} for recent reviews. In this
section we focus on the two operators that induce one-loop corrections
to $\mh^2$ proportional to $g_t^2\,\mt^4/\MS^2$ and two-loop
corrections proportional to $g_t^2\,g_3^2\,\mt^4/\MS^2$\,, i.e.~the
terms suppressed by $\mt^2/\MS^2$\, in what are usually denoted as
one-loop $\oat$ and two-loop $\oatas$ corrections to the Higgs mass,
where $\alpha_t \equiv g_t^2/(4\pi)$ and
$\alpha_s \equiv g_3^2/(4\pi)$. We write the Lagrangian of the SM
extended by dimension-six operators as
\beq
\label{dim6}
{\cal L}_{\scriptscriptstyle {\rm EFT}} ~=~
{\cal L}_{\scriptscriptstyle {\rm SM}} ~-~ c_6\,|H|^6 ~+~ 
\left(\, c_t\,|H|^2 \,\overline{t_\smallR} 
\, H^{\scriptscriptstyle T}\epsilon\,q_\smallL 
~+~ {\rm h.c.}\, \right)~,  
\eeq
where $q_\smallL$ and $t_\smallR$ are third-generation quarks,
$\epsilon$ is the antisymmetric tensor (with $\epsilon_{12}=1$) acting
on the $SU(2)$ indices, and, to fix our notation,
\beq
\label{lagsm}
{\cal L}_{\scriptscriptstyle {\rm SM}} ~\supset~
- m_{\smallH}^2\,|H|^2 ~-~\frac{\lambda}{2}\,|H|^4 ~+~ 
\left(\, g_t\,\overline{t_\smallR} \, H^{\scriptscriptstyle T}\epsilon\,
q_\smallL~+~ {\rm h.c.}\, \right)~.
\eeq

We stress that the choice of considering only the two dimension-six
operators shown in eq.~(\ref{dim6}) implies that our treatment of the
$\ovms$ effects is by no means complete, even when we restrict the
calculation to the ``gaugeless'' limit $g\!=\!g^\prime\!=\!0$. Indeed,
to account for the terms proportional to $g_t^4\,\mt^4/\MS^2$\,, which
are part of the two-loop $\oatq$ corrections to $\mh^2$ already
included in most fixed-order codes, we should include in
eq.~(\ref{dim6}) also dimension-six operators that correct the kinetic
term of the Higgs doublet.\footnote{For those operators several
  definitions are possible. E.g., ref.~\cite{Grzadkowski:2010es} chose
  $(H^{\dagger}H)\Box(H^{\dagger}H)$ and
  $(H^{\dagger}D_\mu \,H)^*(H^{\dagger}D_\mu\, H)$.} Concerning the
resummation of the $\ovms$ logarithmic corrections to $\mh^2$ beyond
two loops, even to account only for the effects controlled by the
highest powers of $g_3$ -- i.e., the $(n\!+\!1)$-loop terms
proportional to $g_t^2\,g_3^{2n}\,\mt^4/\MS^2\,\ln^n(\MS/\mt)$ -- we
should include in eq.~(\ref{dim6}) a set of dimension-six operators
involving gluons.\footnote{Focusing on the CP-even operators, those
  are~ $f^{abc}\,G_{\mu\nu}^a\,G_{\nu\rho}^b\,\,G_{\rho\mu}^c\,$,~
  $|H|^2\,G_{\mu\nu}^a\,G_{\mu\nu}^a$~ and~
  $\overline{t_\smallR} \, \sigma^{\mu\nu}
  \,T^a\,H^{\scriptscriptstyle T}\epsilon\, q_\smallL\,G_{\mu\nu}^a\,$.}
However, it must be kept in mind that the suppression by a factor
$\mt^2/\MS^2$ implies that, for those corrections to be relevant, the
argument of the resummed logarithms cannot be too large. As a result,
there is no guarantee that the three-loop (and higher) logarithmic
effects of $\ovms$ that we could account for via resummation are more
important than other effects that we are neglecting, such as, e.g.,
non-logarithmic three-loop corrections unsuppressed by
$\mt^2/\MS^2$. The sure benefits of extending the SM Lagrangian with
the two dimension-six operators of eq.~(\ref{dim6}) are that $i)$ we
include in our calculation of the Higgs mass the $\ovms$ part of one-
and two-loop corrections that are known to be among the most
significant ones, and $ii)$ we can exploit our knowledge of the size
of those corrections to estimate the theoretical uncertainty
associated to other $\ovms$ effects that we are neglecting.

\bigskip

The boundary conditions on the Wilson coefficients $c_6$ and $c_t$ are
obtained by matching the EFT Lagrangian with the full MSSM Lagrangian
at a renormalization scale $Q\approx\MS$.
We start by remarking that those two coefficients receive
contributions already at the tree level, controlled by the EW gauge
couplings and generated when the heavy Higgs doublet -- whose mass we
denote by $\ma$ -- is integrated out of the MSSM Lagrangian:
\beq
\label{c6cttree}
c_6^{\rm tree}~=~
-\frac{(g^2+g^{\prime\,2})^2}{64\,\ma^2}\,\sin^24\beta~,
~~~~~~~~~~~~~~~
c_t^{\rm tree}~=~
\frac{g_t\,(g^2+g^{\prime\,2})}{8\,\ma^2}\,\sin4\beta\,\cot\beta~.
\eeq
However, in the limit of large $\tan\beta$ both contributions scale
like $1/\tan^2\beta$. For $\tan\beta\gtrsim 10\,$, which we require to
saturate the tree-level prediction for $\mh$ and allow for stop masses
around one TeV, the resulting suppression makes the tree-level
contributions to $c_6$ and $c_t$ numerically comparable with the
one-loop contributions controlled by the EW gauge couplings, which we
are not considering in our study. We will therefore omit the
tree-level contributions of eq.~(\ref{c6cttree}) altogether in what
follows, and we now move on to summarizing our calculation of the one-
and two-loop matching conditions relevant to the $\oat$ and $\oatas$
corrections to the Higgs mass.

\vfill
\paragraph{Matching of $\mathbf{c_t}$\,:}

The one-loop matching condition for $c_t$ can be derived by equating
the expressions for the pole top quark mass computed below and
above the matching scale:
\beq
\label{matchmt}
M_t^{\scriptscriptstyle {\rm pole}} ~=~ 
g_t\,v \,+\, c_t^{1\ell}\,v^3 
\,-\, \Sigma_t^{1\ell}(\mt)^{\scriptscriptstyle {\rm EFT,\msbar}}
~=~ \hat g_t\,\hat v \,-\, 
\Sigma_t^{1\ell}(\mt)^{\scriptscriptstyle {\rm MSSM,\drbar}}~,
\eeq
where $\Sigma_t^{1\ell}(\mt)$ is the one-loop self energy of the top
quark computed with the external momentum $p^2=\mt^2$, and $v$ is the
Higgs vev in the EFT, while $\hat v = \sqrt{v_1^2 + v_2^2}$ is the
corresponding quantity in the MSSM. We adopt as usual the $\drbar$
scheme for the MSSM calculation and the $\msbar$ scheme for the EFT
calculation (note, however, that $c_t^{1\ell}$ is the same in both
schemes).
We focus here on the ${\cal O}(g_3^2)$ and ${\cal O}(g_t^3\,g_3^2)$
contributions to the matching conditions for $g_t$ and $c_t$,
respectively, which are necessary to reproduce the two-loop $\oatas$
corrections to the Higgs mass.  Defining
$\hat g_t(Q) = g_t(Q) \,(1+\Delta g_t^s)$, and considering that the
distinction between $v$ and $\hat v$ does not matter at
${\cal O}(g_3^2)$, we can extract $\Delta g_t^s$ and $c_t^{1\ell}$
from the terms of ${\cal O}(v)$ and ${\cal O}(v^3)$, respectively, in
an expansion of the stop-gluino contribution to the top self energy in
powers of $v$. Starting from eq.~(B2) of ref.~\cite{Degrassi:2001yf}
for the unexpanded self energy, we find
\bea
\label{dgts}
\Delta g_t^{s} &=& - \frac{g_3^2}{(4\pi)^2}\,C_F\,\left[
1 \,+\,\ln\frac{\mg^2}{Q^2} 
\,+\, \wt F_6(\xq)
\,+\, \wt F_6(\xu)
\,-\, \frac{X_t}{\mg}\,\wt F_9(\xq,\xu)\right]~,\\[3mm]
c_t^{1\ell}(Q) &=& \frac{\hat g_t^3\,g_3^2}{(4\pi)^2}\,
\frac{C_F}{\mg^2}
\left\{\, \frac{11 + \xq^2\,(2\,\xq^2-7)}{6\,(\xq^2-1)^3} 
- \frac{2\,\ln \xq}{(\xq^2-1)^4}\right.\nn\\[2mm]
&&~~~~~~~~~~~~~~~~+\left(\frac{X_t}{\mg}-\frac{X_t^2}{2\,\mg^2}\right)
\left[\frac{\xq^2-5}{2\,(\xq^2-1)^2\,(\xu^2-1)}
-\frac{4\,\ln\xq}{(\xq^2-1)^3\,(\xq^2-\xu^2)}\right]\nn\\[2mm]
&&~~~~~~~~~~~~~~~~\left. -\,\frac{2\,X_t^3}{\mg^3}
\left[\frac{1}{(\xq^2-1)\,(\xq^2-\xu^2)^2} 
- \frac{2\,(2\,\xq^4-\xq^2-\xu^2)\,\ln\xq}{(\xq^2-1)^2\,(\xq^2-\xu^2)^3}
\,\right]\,\right\}\nn\\[1mm]
&+&~~\biggr[\,\xq~\longleftrightarrow~\xu\,\biggr]~,
\label{ctfull}
\eea
where the functions $\wt F_6$ and $\wt F_9$ can be found in appendix A
of ref.~\cite{Bagnaschi:2014rsa}, we defined $\xq = m_{Q_3}/\mg$
and $\xu = m_{U_3}/\mg$\,, and the term in the last line of
eq.~(\ref{ctfull}) is obtained from the terms in the first three lines
by swapping $\xq$ and $\xu$. We note that the right-hand side of
eq.~(\ref{ctfull}) does not depend explicitly on the scale $Q$. For
the simplified choice $m_{Q_3}=m_{U_3}=\mg=\MS$, the
${\cal O}(g_t^3\,g_3^2)$ contribution to the matching condition for
$c_t$ reduces to
\beq
c_t^{1\ell}(Q) ~=~ \frac{\hat g_t^3\,g_3^2}{(4\pi)^2}\,
\frac{C_F}{12\,\MS^2}
\left( 6 ~+~ 6\,\frac{X_t}{\MS} - 3\,\frac{X_t^2}{\MS^2} 
- 2\,\frac{X_t^3}{\MS^3}\right)~. 
\label{ctlim}
\eeq

\paragraph{Matching of $\mathbf{c_6}$\,:}

The matching condition for the Wilson coefficient of the operator
$|H|^6$ in eq.~(\ref{dim6}) can, in analogy with the calculation of
the matching condition for the quartic Higgs coupling described in
section~\ref{sec:lambda2loop}, be obtained from the derivatives with
respect to the Higgs field of the sfermion contributions to the
effective potential of the MSSM. In particular, the ${\cal O}(g_t^6)$
contribution to the one-loop coefficient $c_6^{1\ell}$ at the matching
scale $Q$ reads
\beq
c_6^{1\ell}(Q) ~=~ \frac{1}{36}\,
\left.\frac{\partial^6 \Delta V^{1\ell,\,\tilde t}}
{\partial^3H^\dagger\partial^3 H}\,\right|_{H=0},
\eeq
where $\Delta V^{1\ell,\,\tilde t}$ is the stop contribution to the
Coleman-Weinberg potential of the MSSM
\beq
\Delta V^{1\ell,\,\tilde t}~=~\frac{N_c}{(4\pi)^2}\,\sum_{i=1,2}\,
\frac{m_{\tilde t_i}^4}{2}\,\left(\ln\frac{\ti}{Q^2} - \frac32\right)~.
\eeq
As outlined in section 2.3 of ref.~\cite{Bagnaschi:2014rsa}, the
derivatives of $\Delta V^{1\ell,\,\tilde t}$ with respect to the Higgs
field can be easily computed after expressing the stop masses $\ti$ as
functions of the field-dependent top mass $\mt = \hat g_t\,|H|$,
leading to
\beq
\label{chain}
\left.\frac{\partial^6 \Delta V^{1\ell,\,\tilde t}}
{\partial^3H^\dagger\partial^3 H}\,\right|_{H=0}=~
\hat g_t^6\,\left[6\,V^{(3)}_{ttt}\,+\,18\,\mt^2\,V^{(4)}_{tttt}\,+\,
9\,\mt^4\,V^{(5)}_{ttttt}\,+\,\mt^6\,V^{(6)}_{tttttt}
\right]_{\mt\rightarrow 0}~,
\eeq
where we used for the derivatives of the one-loop potential shortcuts
analogous to those defined in eq.~(\ref{shortcut}) for the derivatives
of the two-loop potential. Explicitly, we find
\bea
c_6^{1\ell}(Q) &=&
\frac{\hat g_t^6}{(4\pi)^2}\,\frac{N_c}{m_{Q_3}m_{U_3}}\,
\left\{\frac{1+x_t^2}{6\,x_t}~-~\frac{\wt X_t}{2}
~+~\wt X_t^2\,\left[\,\frac{x_t\,(1+x_t^2)}{2\,(1-x_t^2)^2}
\,+\,\frac{2\,x_t^3\,\ln x_t}{(1-x_t^2)^3}\,\right]\right.\nn\\[2mm]
&&~~~~~~~~~~~~~~~~~~~~~~~\left.
-\,\wt X_t^3\,\left[\,\frac{x_t^2\,(1+10\,x_t^2+x_t^4)}{6\,(1-x_t^2)^4}
\,+\,\frac{2\,x_t^4\,(1+x_t^2)\,\ln x_t}{(1-x_t^2)^5}\,\right]\right\}~,
\label{c61loop}
\eea
where, following the notation of eq.~(\ref{oneloop}), we defined
$x_t = m_{Q_3}/m_{U_3}$ and $\wt X_t = X_t^2/ (m_{Q_3}m_{U_3})$.
Eq.~(\ref{c61loop}) agrees with the corresponding results in
refs.~\cite{Huo:2015nka, Drozd:2015rsp}\,\footnote{In
  ref.~\cite{Drozd:2015rsp} there is a misprint in the last line of
  eq.~(D.4): the logarithmic term should come with a minus sign.},
which employed the method known as ``covariant derivative
expansion''~\cite{Gaillard:1985uh, Cheyette:1987qz, Henning:2014wua}
to compute the one-loop matching conditions for all bosonic
dimension-six operators induced by integrating the squarks out of the
MSSM Lagrangian. In the limit of degenerate squark masses
$m_{Q_3}=m_{U_3}=\MS$, the ${\cal O}(g_t^6)$ contribution to
$c_6^{1\ell}$ reduces to
\beq
\label{c61loopsimpl}
c_6^{1\ell}(Q) ~=~
\frac{\hat g_t^6}{(4\pi)^2}\,\frac{N_c}{\MS^2}\,
\left(\frac13 \,-\,\frac{X_t^2}{2\,\MS^2}  \,+\, \frac{X_t^4}{6\,\MS^4}
\,-\,\frac{X_t^6}{60\,\MS^6}\,\right)~,
\eeq
in agreement with the result presented long ago in
ref.~\cite{Casas:1994us}. 

\bigskip

The ${\cal O}(g_t^6g_3^2)$ contribution to the two-loop coefficient
$c_6^{2\ell}$ at the matching scale $Q$ reads
\beq
\label{c62loop}
c_6^{2\ell}(Q) ~=~ \frac{1}{36}\,
\left.\frac{\partial^6 \Delta V^{2\ell,\,\tilde t}}
{\partial^3H^\dagger\partial^3 H}\,\right|_{H=0}
-~\delta c_6^{\scriptscriptstyle {\rm EFT}}
~+~\delta c_6^{{\rm shift},\, \tilde t}~,
\eeq
where $\Delta V^{2\ell,\,\tilde t}$ denotes the contribution to the
MSSM scalar potential from two-loop diagrams involving the strong
gauge interactions of the stop squarks, given e.g.~in eq.~(28) of
ref.~\cite{Bagnaschi:2014rsa}. The derivatives of
$\Delta V^{2\ell,\,\tilde t}$ are again obtained from
eq.~(\ref{chain}) after expressing the stop masses and mixing angle as
functions of the field-dependent top mass. As in the case of the
two-loop matching condition for the quartic Higgs coupling discussed
in section \ref{sec:lambda2loop}, in the derivatives of the two-loop
scalar potential we find terms proportional to $\ln(\mt^2/Q^2)$, which
would diverge in the limit of vanishing top mass. Those terms,
however, cancel against analogous terms in
$\delta c_6^{\scriptscriptstyle {\rm EFT}}$, which represents the
one-loop contribution of the dimension-six operators to the Wilson
coefficient of $|H|^6$ as computed in the EFT. In particular, the
contribution relevant at ${\cal O}(g_t^6g_3^2)$ arises from a box
diagram with a top-quark loop, three regular Yukawa vertices and one
dimension-six vertex. We find
\beq
\delta c_6^{\scriptscriptstyle {\rm EFT}}~=~
-\frac{N_c}{(4\pi)^2}\,g_t^3\,c_t^{1\ell}(Q)
\left(4\,\ln\frac{\mt^2}{Q^2}+\frac{32}3\right)~,
\eeq
where for $c_t^{1\ell}(Q)$ we use the ${\cal O}(g_t^3\,g_3^2)$
contribution to the matching condition given in
eq.~(\ref{ctfull}). Finally, the third term on the right-hand side of
eq.~(\ref{c62loop}) arises from the fact that, in analogy with our
two-loop calculation of the quartic Higgs coupling, we choose to
express the one-loop stop contribution to $c_6^{1\ell}$ in terms of
the $\msbar$-renormalized top Yukawa coupling of the EFT, i.e.~$g_t$,
as opposed to the $\drbar$-renormalized coupling of the MSSM, i.e.~the
$\hat g_t$ entering
eqs.~(\ref{chain})--(\ref{c61loopsimpl})\,\footnote{On the other hand,
  in the two-loop corrections the distinction between $g_t$ and
  $\hat g_t$ amounts to a higher-order effect.}. The resulting
${\cal O}(g_t^6\,g_3^2)$ shift in $c_6^{2\ell}$ reads
\beq
\label{dc6shift}
\delta c_6^{{\rm shift},\,\tilde t} ~=~ 6\,\Delta g_t^s\,c_6^{1\ell}(Q)~,
\eeq
where $\Delta g_t^s$ is given in eq.~(\ref{dgts}) and $c_6^{1\ell}(Q)$
is given in eq.~(\ref{c61loop}).

\bigskip

The analytic formula for $c_6^{2\ell}(Q)$ for generic stop and gluino
masses is too lengthy to be printed, and we make it available on
request in electronic form. For the simplified choice
$m_{Q_3}=m_{U_3}=\mg=\MS$, we obtain
\bea
c_6^{2\ell}(Q) &=&
-\frac{\hat g_t^6\,g_3^2}{(4\pi)^4}\,\frac{C_FN_c}{\MS^2}
\left[\frac23-\frac{4\,X_t}{\MS}+ \frac{X_t^2}{\MS^2}
+\frac{14\,X_t^3}{3\,\MS^3}+\frac{X_t^4}{6\,\MS^4}
-\frac{13\,X_t^5}{10\,\MS^5}-\frac{19\,X_t^6}{180\,\MS^6}
+\frac{X_t^7}{10\,\MS^7}\right.\nn\\[2mm]
&&~~~~~~~~~~~~~~~~~~~~\left.
+\left(\frac83+\frac{2\,X_t}{\MS}- \frac{4\,X_t^2}{\MS^2}
-\frac{2\,X_t^3}{\MS^3}+\frac{X_t^4}{\MS^4}
+\frac{2\,X_t^5}{5\,\MS^5}-\frac{X_t^6}{30\,\MS^6}\right)
\ln\frac{\MS^2}{Q^2}\,\right]~.\nn\\
\eea
We also remark that $c_6^{2\ell}(Q)$ contains terms enhanced by powers
of the ratios between the gluino mass and the stop masses. In
particular, in the simplified scenario where $m_{Q_3}=m_{U_3}=\MS$,
$\,X_t= \pm \sqrt{6}\,\MS$ and $\mg \gg \MS$ we find 
\bea
c_6^{2\ell}(Q) &=&
-\frac{\hat g_t^6\,g_3^2}{(4\pi)^4}\,\frac{4\,C_FN_c}{15\,\MS^2}
\left[1 \,-\, 18\,\ln\frac{\MS^2}{Q^2}
\,+\, 13\,\ln\frac{\mg^2}{\MS^2} \,-\, \frac{\mg}{\MS}\left(
\pm \, 9 \,\sqrt{6} \,
+\, 31\,\frac{\mg}{\MS}\right)\!\left(1-\ln\frac{\mg^2}{Q^2}
\right)\right.\nn\\
&&~~~~~~~~~~~~~~~~~~~~~
\left.~+~{\cal O}\left(\frac{\MS}{\mg}\right)\,\right]~,
\eea
where the sign of the first term within round brackets corresponds to
the sign of $X_t$. The presence of power-enhanced terms in the
heavy-gluino limit is a well-known consequence of the $\drbar$
renormalization of the parameters in the stop sector, as discussed in
ref.~\cite{Degrassi:2001yf} for the fixed-order calculation of the
MSSM Higgs masses and in ref.~\cite{Vega:2015fna} for the EFT
calculation. Those terms would be removed from the two-loop part of
$c_6$ if we interpreted the soft SUSY-breaking stop masses $m_{Q_3}$
and $m_{U_3}$ and the stop mixing $X_t$ entering the one-loop part as
``on-shell''-renormalized parameters.

\paragraph{Comparison with the fixed-order calculation of $\mathbf{\mh^2}$\,:}

We now discuss how the inclusion in the EFT Lagrangian of the
dimension-six operators shown in eq.~(\ref{dim6}) allows us to
reproduce the $\omtms$ terms in the $\oat$ and $\oatas$ corrections to
$\mh^2\,$.  Expanding the neutral component of the Higgs doublet as
$H^0 = v + (h\,+\,i\,G)/\sqrt{2}\,$, and exploiting the minimum
condition of the scalar potential to remove the mass parameter
$m_{\smallH}^2$, we can write the Higgs-boson mass as
\beq
\label{mhEFT}
\mh^2 ~=~ 2\,\lambda\,v^2 ~+~ 12\, c_6\,v^4 
~+\, \dmhq~,
\eeq
where $\dmhq$ contains the radiative corrections to the tree-level
prediction for the Higgs mass, as computed in the EFT. To avoid the
occurrence of large logarithms in these corrections, the couplings
$\lambda$ and $c_6$ in eq.~(\ref{mhEFT}) should be computed at a
renormalization scale $\qew$ of the order of the masses of the
particles running in the loops. Focusing on the one- and two-loop
terms that account for the desired $\oat$ and $\oatas$ corrections, we
find
\bea
\dmhq&=&
\frac{N_c}{(4\pi)^2}\,\left[
-4\,g_t^4\,v^2\,\ln\frac{g_t^2\,v^2}{\qew^2}
+8\,g_t\,c_t\,v^2\,\mt^2\,\left(1-6\,\ln\frac{\mt^2}{\qew^2}\right)\,
\right]\nn\\[2mm]
&+& \frac{C_F N_c}{(4\pi)^4}\,8\,g_t^2\,g_3^2\,\mt^2
\left(3\,\ln^2\frac{\mt^2}{\qew^2}~+~\ln\frac{\mt^2}{\qew^2}\right)~,
\label{dmhEFT}
\eea
where the first line is the contribution of one-loop diagrams
involving top quarks, the second line is the contribution of two-loop
diagrams involving top quarks and gluons computed in the $\msbar$
scheme (the latter was given, e.g., in ref.\cite{Degrassi:2012ry}),
and we can typically take $\qew\approx\mt$.
Since for the purpose of this calculation the coefficient $c_t$ is
first generated at one loop, in the first line of eq.~(\ref{dmhEFT})
we have exploited the relation $\mt = g_t\,v + c_t\,v^3$ and
retained\,\footnote{Before the expansion in $c_t$ the one-loop
  contribution to $\Delta\mh^2$ involving top quarks reads, in our
  EFT,
\beq
\nn
(\dmhq)^{1\ell,\,t} ~=~ 
\frac{N_c}{(4\pi)^2}\,\left[\,
2\,\mt^2\,\left(g_t+3\,c_t\,v^2\right)^2
\left(1-3\,\ln\frac{\mt^2}{\qew^2}\right)
\,-\, 2\,\frac{\mt^3}{v}\,\left(g_t-3\,c_t\,v^2\right)
\left(1-\ln\frac{\mt^2}{\qew^2}
\right)\,\right]~.
\eeq}
only terms linear in $c_t$ (note that in those
terms, as well as in those of the second line, the difference between
$\mt^2$ and $g_t^2\,v^2$ amounts to a higher-order effect).
Collecting all the terms in eq.~(\ref{mhEFT}) that involve the
coefficients of dimension-six operators, we thus find for the one- and
two-loop $\omtms$ terms
\bea
\label{oatdim6}
\left(\mh^2\right)_{\scriptscriptstyle {\rm dim6}}^{\oat} &=& 
12\,v^4 \,c_6^{1\ell}(Q)~,\\[4mm]
\label{oatasdim6}
\left(\mh^2\right)_{\scriptscriptstyle {\rm dim6}}^{\oatas} &=& 
12\,v^4 \left[c_6^{2\ell}(Q)
~+ \left.\frac{d c_6}{d \ln Q^2}\right|_{g_t^3c_t}
\! \ln\frac{\qew^2}{Q^2}\right]
+\, \frac{N_c}{(4\pi)^2}\,8\,g_t\,c_t\,v^2\,\mt^2\,
\left(1-6\,\ln\frac{\mt^2}{\qew^2}\right)~,\nn\\
\eea
where the one-loop beta function of $c_6$ entering the squared
brackets in eq.~(\ref{oatasdim6}) accounts, at the two-loop level, for
the fact that in eq.~(\ref{mhEFT}) the coefficient $c_6$ should be
computed at the low scale $\qew$.  Isolating the relevant terms in the
RGEs for the dimension-six operators given in
refs.~\cite{Elias-Miro:2013mua, Jenkins:2013zja, Jenkins:2013wua,
  Alonso:2013hga}, we have
\beq
\label{RGEdim6}
\frac{d c_6}{d \ln Q^2}~\supset~
-4\,N_c\,\frac{g_t^3\,c_t}{(4\pi)^2}~,~~~~~~~~
\frac{d c_t}{d \ln Q^2}~\supset~
-3\,C_F\,\frac{g_3^2\,c_t}{(4\pi)^2}~.
\eeq
However, for the coefficient $c_t$ entering eq.~(\ref{oatasdim6}) we
can use directly the value obtained at the matching scale, see
eq.~(\ref{ctfull}), because its scale dependence amounts to a
three-loop effect in $\mh^2$. We thus obtain
\beq
\label{oatasRGE}
\left(\mh^2\right)_{\scriptscriptstyle {\rm dim6}}^{\oatas} ~=~
 12\, v^4\, c_6^{2\ell}(Q) 
~+~ \frac{N_c}{(4\pi)^2}\,8\,g_t\,c_t^{1\ell}(Q)\,v^2\,\mt^2\,
\left(1-6\,\ln\frac{\mt^2}{Q^2}\right)~.
\eeq

Expanding in powers of $\mt^2$ the analytic results of
ref.~\cite{Degrassi:2001yf} for the $\oat$ and $\oatas$ corrections to
$\mh^2$ in the MSSM, we checked that eqs.~(\ref{oatdim6}) and
(\ref{oatasRGE}) do indeed reproduce the one- and two-loop $\omtms$
terms of those corrections, respectively. To this purpose, it is
necessary to take into account that in ref.~\cite{Degrassi:2001yf} the
top mass and Yukawa coupling entering the one-loop part of the
corrections to $\mh^2$ are assumed to be MSSM parameters renormalized
in the $\drbar$ scheme, whereas, as discussed earlier, we choose to
express $c_6^{1\ell}(Q)$ in terms of the EFT coupling $g_t$
renormalized in the $\msbar$ scheme. To perform the comparison with
the fixed-order calculation of $\mh^2$ we must therefore omit the term
$\delta c_6^{{\rm shift},\,\tilde t}$ in our formula for
$c_6^{2\ell}(Q)$, see eqs.~(\ref{c62loop}) and (\ref{dc6shift}).


\subsection{Impact of dimension-six operators on the Higgs mass
  prediction}

In this section we illustrate the numerical impact of the
dimension-six operators of eq.~(\ref{dim6}) on the EFT prediction for
the Higgs mass. We modified the code \hssusy~\cite{HSSUSY},
implementing the matching conditions for $c_6$ and $c_t$ at the SUSY
scale, their evolution down to the EW scale through the RGEs of
eq.~(\ref{RGEdim6})\footnote{Note that we neglect additional terms in
  those RGEs, as well as the contributions of the dimension-six
  operators to the RGEs of the SM couplings~\cite{Jenkins:2013zja},
  because they do not contribute to the $\oat$ and $\oatas$
  corrections to $\mh^2$.}, and their effects at the EW scale, both on
the calculation of $\mh^2$ -- see eqs.~(\ref{mhEFT}) and
(\ref{dmhEFT}) -- and on the determination of the top Yukawa
coupling. In particular, the latter becomes
\beq
\label{matchgt}
g_t(\qew) ~=~ \frac{\overline m_t}{v}\,-\, c_t\,v^2~,
\eeq
where $\overline m_t$ denotes the $\msbar$-renormalized top mass,
extracted at the scale $\qew$ from
$M_t^{\scriptscriptstyle {\rm pole}}$ with SM formulae, and we neglect
the effects of dimension-six operators that do not contribute at
${\cal O}(g_3^2)$.
We note that the $c_t$-induced shift on the matching condition for the
top Yukawa coupling, eq.~(\ref{matchgt}) above, affects all
corrections controlled by $g_t$ to the quartic Higgs coupling --
namely, the threshold corrections at the SUSY scale and the
renormalization-group evolution down to the EW scale -- as well as the
top-quark contributions to $\Delta\mh^2$ given in
eq.~(\ref{dmhEFT}). This results in an ``indirect'' contribution of a
dimension-six operator to the EFT prediction for $\mh^2$, which
combines with the ``direct'' contributions controlled by $c_6$ and
$c_t$ in eqs.~(\ref{mhEFT}) and (\ref{dmhEFT}).

\bigskip

\begin{figure}[t]
\begin{center}
\includegraphics[width=0.48\textwidth]{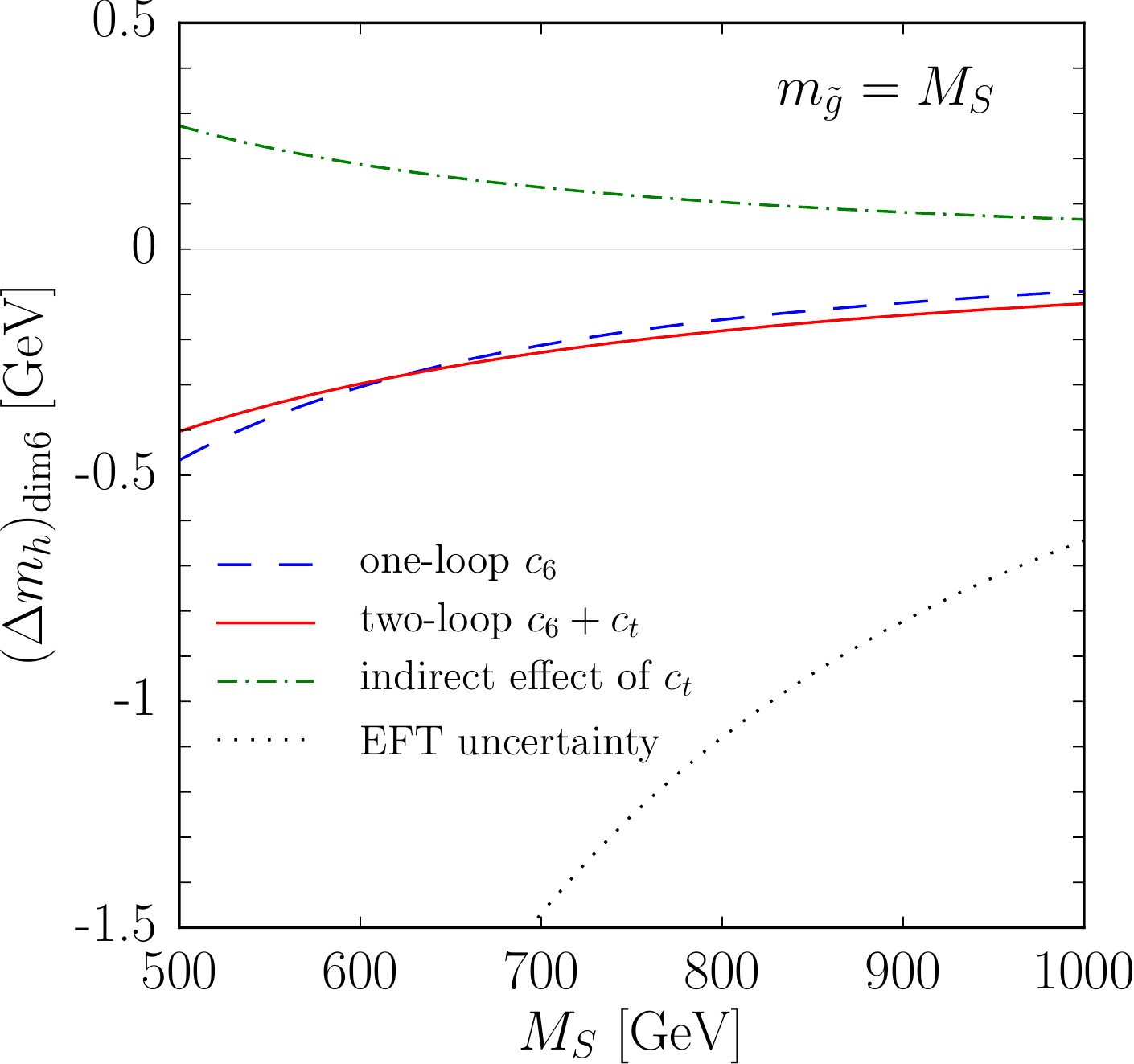}~~~~~
\includegraphics[width=0.48\textwidth]{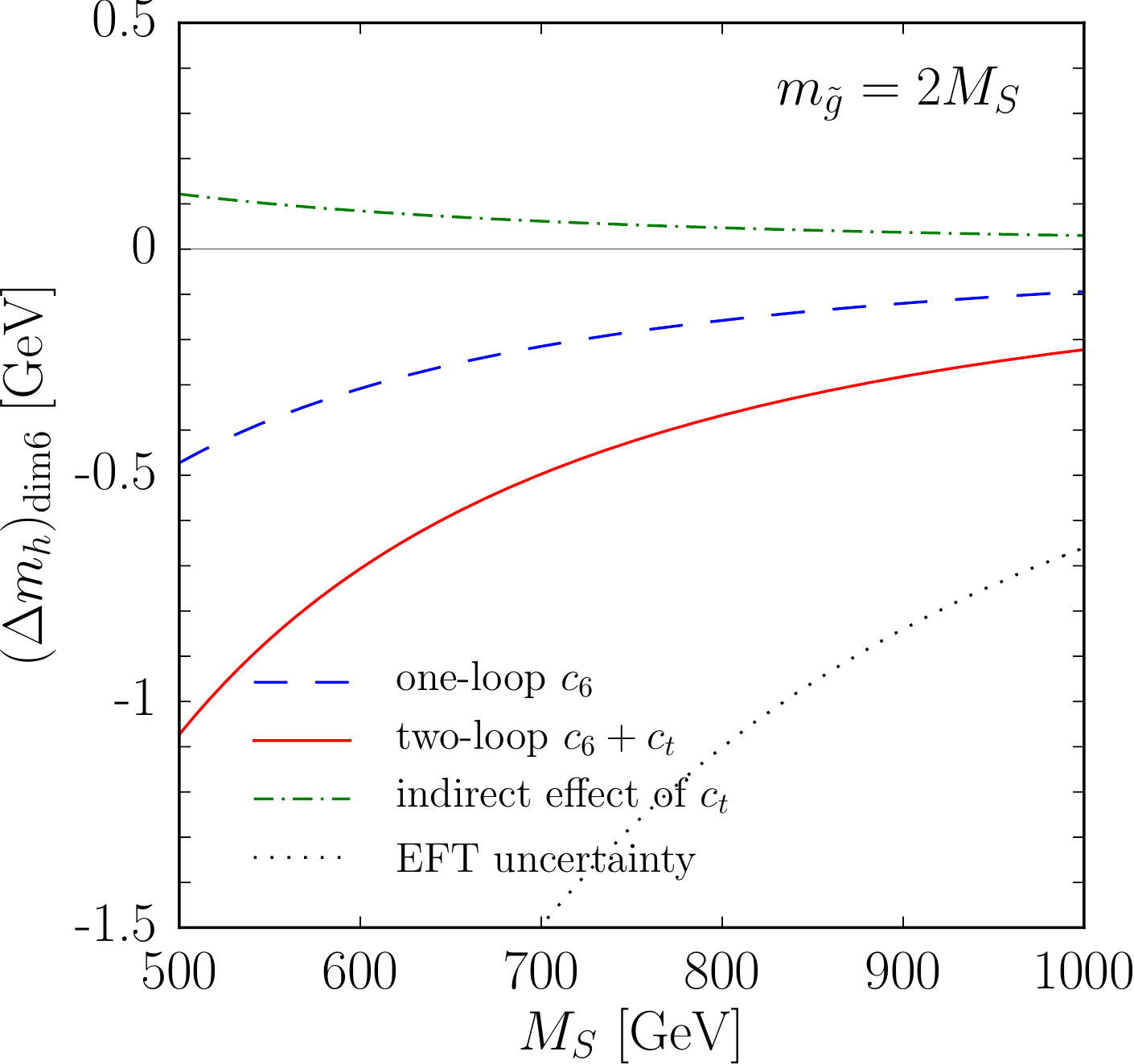}
%
\caption{\em Effects of dimension-six operators on the EFT prediction
  for the Higgs mass, as a function of a common stop mass scale $\MS$,
  for $X_t = \sqrt{6}\,\MS$. In the left plot we take $\mg=\MS$,
  whereas in the right plot we take $\mg=2\,\MS$. The meaning of the
  different curves and the values of the remaining MSSM parameters are
  described in the text.}
  \label{fig:dim6ms}
\end{center}
\end{figure}

In figure~\ref{fig:dim6ms} we show the deviation induced in the EFT
prediction for the Higgs mass by the presence of the dimension-six
operators of eq.~(\ref{dim6}).
The SM parameters used as input for \hssusy\ are the same as those
listed at the beginning of section~\ref{sec:tlnumbers}. We consider a
simplified MSSM scenario with $\tan\beta=20$ and all soft
SUSY-breaking masses of sfermions and EW gauginos, as well as the
heavy Higgs-doublet mass $\ma$ and the higgsino mass $\mu$, set equal
to a common SUSY scale $\MS$; the trilinear Higgs-stop coupling $A_t$
is fixed by the maximal mixing condition
$A_t-\mu\,\cot\beta = \sqrt{6}\,\MS\,$, and $A_b=A_{\tau}=A_t$;
finally, the gluino mass is set to $\mg = \MS$ in the left plot and to
$\mg = 2\,\MS$ in the right plot.  We vary the common SUSY scale
between $\MS= 500$~GeV and $\MS= 1$~TeV, and interpret the soft
SUSY-breaking stop masses and $A_t$ as $\drbar$-renormalized
parameters at the matching scale $Q=\MS$.  We remark that, in the
considered range of $\MS$, the prediction of \hssusy\ for $\mh$
(before the introduction of the dimension-six operators) varies
between $120.2$~GeV and $123$~GeV in the left plot and between
$118.7$~GeV and $121.9$~GeV in the right plot, always several GeV
below the value measured at the LHC. Therefore, rather than depicting
fully realistic scenarios, the figure is meant to illustrate the
relative importance of the different effects induced by dimension-six
operators, and how those effects get suppressed by an increase in the
SUSY scale.

The dashed blue lines in the plots of figure~\ref{fig:dim6ms}
represent the inclusion of the sole operator $|H|^6$, with the
coefficient $c_6$ computed at one loop and ``frozen'' at the matching
scale $Q=\MS$. This accounts for the $\omtms$ part of the $\oat$
corrections to the Higgs mass, as given in eq.~(\ref{oatdim6}). We see
that, in these scenarios, the corresponding shift in $\mh$ is negative
and rather modest, decreasing from about $470$~MeV for $\MS=500$~GeV
to about $90$~MeV for $\MS=1$~TeV.

The solid red lines in the plots of figure~\ref{fig:dim6ms} represent
instead the inclusion of both of the operators of eq.~(\ref{dim6}),
with coefficients $c_6$ and $c_t$ computed at two loops and one loop,
respectively, and evolved between the scales $\MS$ and $\qew$ with the
RGEs of eq.~(\ref{RGEdim6}). This accounts also for the $\omtms$ part
of the $\oatas$ corrections to the Higgs mass, as given in
eq.~(\ref{oatasRGE}). In the scenario shown in the left plot,
corresponding to $\mg = \MS$, the two-loop $\omtms$ corrections appear
to be rather small, never reaching even $\pm 100$~MeV in the
considered range of $\MS$. However, we must take into account that the
difference between the dashed blue and solid red lines results from
the combination of several effects, namely: $i)$ the $c_t$-induced
shift in the value of $g_t$ used in the whole calculation, see
eq.~(\ref{matchgt}), whose ``indirect'' effects on $\mh$ we show for
illustration as the dot-dashed green lines; $ii)$ the inclusion of the
two-loop part of the matching condition for $c_6$ at the SUSY scale;
$iii)$ the evolution of $c_6$ (and of $c_t$) between the SUSY scale
and the EW scale; $iv)$ the terms controlled by $c_t$ in the radiative
corrections to the Higgs mass at the EW scale, see
eq.~(\ref{dmhEFT}). In the scenario of the left plot, the first three
of these effects shift $\mh$ by several hundred MeV each for
$\MS=500$~GeV, but they undergo significant cancellations, whereas the
fourth effect is considerably less important. On the other hand, in
the scenario shown in the right plot, corresponding to $\mg = 2\,\MS$,
the ``indirect'' effects of the shift in $g_t$ are reduced due to a
smaller value of $c_t$, and the two-loop contribution to the matching
of $c_6$ at the SUSY scale doubles in size and changes sign, with the
result that the combined effects of the two-loop $\omtms$ corrections
are much more significant than in the left plot, further decreasing
the prediction for $\mh$ by about $600$~MeV for $\MS=500$~GeV and
about $130$~MeV for $\MS=1$~TeV.

To assess the relevance of the $\omtms$ logarithmic effects beyond two
loops, we removed from the EFT prediction for the Higgs mass the
higher-order terms that are picked up by solving numerically the RGEs
of the Wilson coefficients in eq.~(\ref{RGEdim6}). In practice, we
compared our results for $\mh$ with those obtained by ``freezing''
$c_t$ at the SUSY scale and truncating the evolution of $c_6$ to the
first order in the perturbative expansion -- see the terms within
square brackets in eq.~(\ref{oatasdim6}). We found that these
higher-order logarithmic effects are very small in the considered
scenarios: even in the one with $\mg=\MS$, characterized by a larger
value of $c_t$ and hence a stronger scale dependence of both $c_t$ and
$c_6$, the resulting shift in $\mh$ reaches a maximum of about
$20$~MeV for $\MS \approx 600$~GeV, then decreases for larger $\MS$ as
the suppression by a factor $\mt^2/\MS^2$ begins to prevail over the
logarithmic enhancement.
 
Finally, the dotted black lines in the plots of
figure~\ref{fig:dim6ms} represent a naive estimate of the overall size
of the $\ovms$ corrections to the Higgs mass, corresponding to the
``EFT uncertainty'' implemented in the code $\SHD$. Following
ref.~\cite{Vega:2015fna}, we obtain that estimate by multiplying the
contribution to $\Delta \lambda^{1\ell}$ from each SUSY particle with
mass $M_i$ by a factor\,\footnote{To be conservative, we adjust the
  signs in the rescaling factors for scalars and EW-inos so that the
  resulting shifts in $\Delta \lambda^{1\ell}$ add up. The upper edge
  of the uncertainty band, not shown in the plots, can be obtained by
  reversing all signs.}  $(1 \pm 2\, v^2/M_i^2)$. It appears that,
even in the scenario of the right plot where the computed $\omtms$
corrections to the Higgs mass are more significant, the $\SHD$
estimate of those effects is larger by about a factor of
three. Therefore, even if the one- and two-loop $\ovms$ corrections to
$\mh^2$ that are not included in our analysis -- such as, e.g., the
two-loop corrections proportional to $g_t^4\,m_t^4/\MS^2$ -- were as
large as the ones that we did compute and had the same sign, the
estimate of the ``EFT uncertainty'' implemented in $\SHD$ would turn
out to be sufficiently conservative in the considered scenarios.

\bigskip

\begin{figure}[t]
\begin{center}
\includegraphics[width=0.48\textwidth]{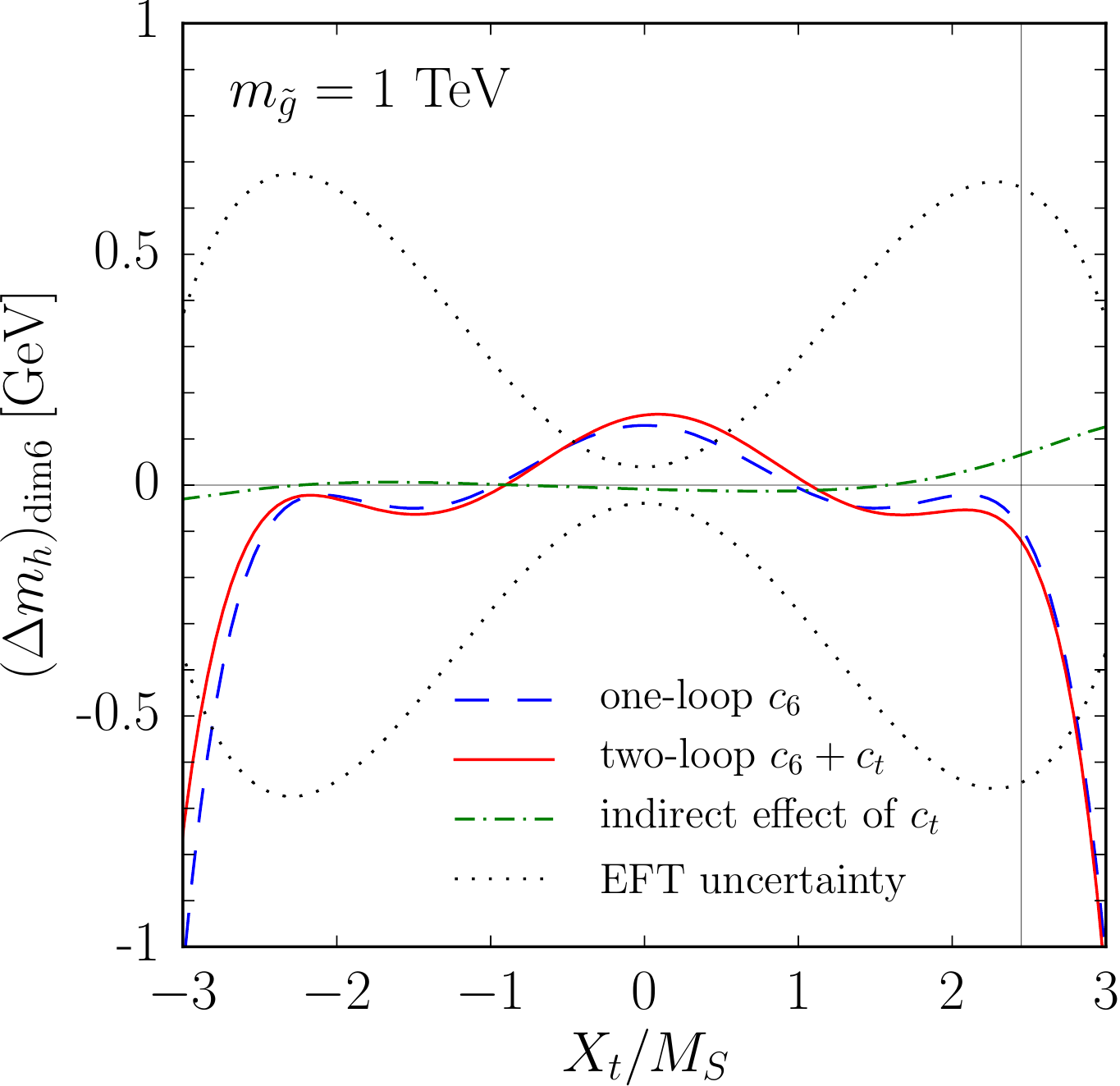}~~~~~
\includegraphics[width=0.48\textwidth]{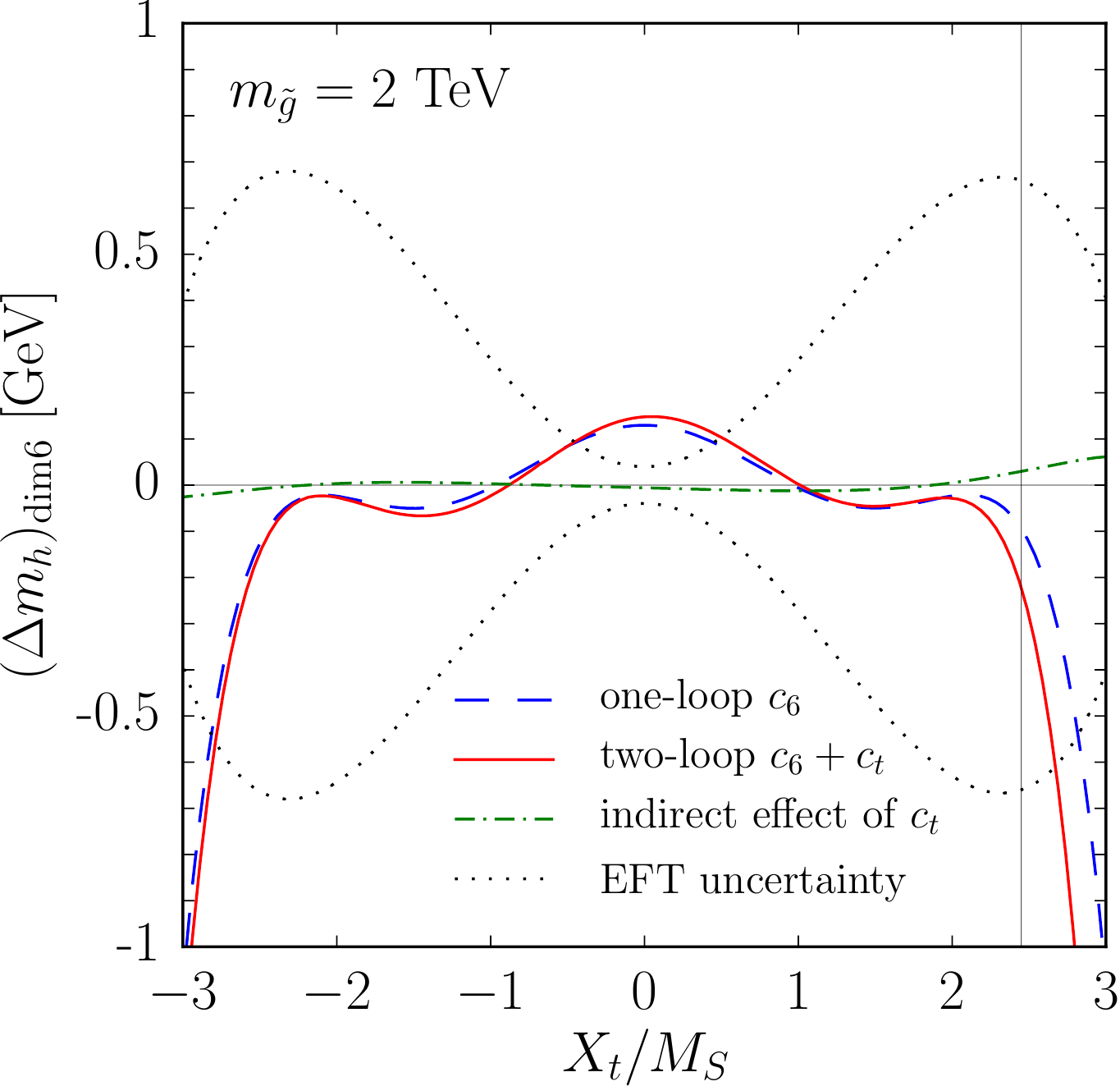}
%
\caption{\em Effects of dimension-six operators on the EFT prediction
  for the Higgs mass, as a function of the ratio $X_t/\MS$, for a
  common stop mass scale $\MS = 1~{\rm TeV}$. In the left plot we take
  $\mg=1~{\rm TeV}$, whereas in the right plot we take
  $\mg=2~{\rm TeV}$. The thin vertical lines in the two plots mark the
  condition $X_t/\MS = \sqrt{6}\,$. The meaning of the different
  curves and the values of the remaining MSSM parameters are the same
  as in figure~\ref{fig:dim6ms}, as described in the text.}
  \label{fig:dim6Xt}
\end{center}
\end{figure}

It is legitimate to wonder whether the relatively small size of the
$\omtms$ corrections found in the scenarios of figure~\ref{fig:dim6ms}
is just an accident, perhaps related to the choice
$X_t = \sqrt{6}\,\MS$ made to ensure a near-maximal prediction for the
Higgs mass. To answer this question, in figure~\ref{fig:dim6Xt} we
show again the deviation induced in the EFT prediction for the Higgs
mass by the presence of the dimension-six operators of
eq.~(\ref{dim6}), this time as a function of the ratio $X_t/\MS$. We
set $\MS=1$~TeV, and take all of the remaining MSSM parameters as in
the two scenarios of figure~\ref{fig:dim6ms}.  In particular, we take
$\mg=1$~TeV in the left plot and $\mg=2$~TeV in the right plot. The
thin vertical lines in the two plots of figure~\ref{fig:dim6Xt} mark
the condition $X_t/\MS = \sqrt{6}$, i.e.~they map the right edge of
the corresponding plots in figure~\ref{fig:dim6ms}.  The meaning of
all other lines is the same as in figure~\ref{fig:dim6ms}.

Figure~\ref{fig:dim6Xt} shows that, for a given value of $\MS$, the
impact of the $\omtms$ corrections to the Higgs mass can indeed be
larger than the one found when $X_t/\MS = \sqrt{6}\,$. This happens in
particular for $X_t \approx 0$, or for values of $|X_t/\MS|$ larger
than $\sqrt{6}$. The figure also shows that for $X_t \approx 0$ the
$\SHD$ estimate of the ``EFT uncertainty'' falls short of the computed
$\omtms$ corrections to the Higgs mass. Indeed, the main contribution
to the $\SHD$ estimate is the one from stops, which -- being
proportional to the corresponding contribution to
$\Delta \lambda^{1\ell}$, see eq.~(\ref{oneloop}) -- is maximized for
$|X_t/\MS| = \sqrt{6}\,$ and vanishes for $X_t=0$ (the small non-zero
value of the ``EFT uncertainty'' visible in the plots at $X_t=0$ is
due to the contributions of EW gauginos and higgsinos). In contrast,
eq.~(\ref{c61loopsimpl}) shows that the one-loop stop contribution to
$c_6$ does not vanish for $X_t=0$, yielding a shift in $\mh$ of about
$130$~MeV.
However, we must recall that moving away from the ``maximal mixing''
condition on $X_t$ results in a significant decrease in the EFT
prediction for the Higgs mass, e.g.~for $X_t = 0$ we would find
$\mh \approx 110.8$~GeV. In order to recover a prediction for $\mh$
within a few GeV from the observed value, we would need to raise the
SUSY scale $\MS$ to several TeV, strongly suppressing all effects of
dimension-six operators. Therefore, the $\SHD$ estimate of the ``EFT
uncertainty'' happens to be at its most conservative precisely in the
region of the MSSM parameter space where the $\omtms$ effects
discussed in this section have a chance to be numerically relevant.

\section{Conclusions}
\label{sec:conclusions}

If the MSSM is realized in nature, both the measured value of the
Higgs mass and the (so-far) negative results of the searches for
superparticles at the LHC suggest some degree of separation between
the SUSY scale $\MS$ and the EW scale. In this scenario the MSSM
prediction for the Higgs mass is subject to potentially large
logarithmic corrections, which can be resummed to all orders in an EFT
approach. Over the past few years this has stimulated a considerable
amount of activity, aimed, on one hand, at refining the EFT
calculation of the MSSM Higgs mass~\cite{Draper:2013oza,
  Bagnaschi:2014rsa, Vega:2015fna}, and, on the other hand, at
combining it with the fixed-order calculations implemented in public
codes for the determination of the MSSM mass
spectrum~\cite{Hahn:2013ria, Bahl:2016brp, Athron:2016fuq,
  Staub:2017jnp}. Here we contributed to these efforts by providing a
complete determination of the two-loop threshold corrections to the
quartic Higgs coupling in the limit of vanishing EW gauge (and
first-two-generation Yukawa) couplings, for generic values of all the
relevant SUSY-breaking parameters. We also studied a class of one- and
two-loop corrections to the Higgs mass suppressed by $\mt^2/\MS^2\,$,
extending the SM Lagrangian with appropriate dimension-six operators.
All of our results are available upon request in electronic form, and
they were also implemented in modified versions of the codes
\SHD~\cite{SusyHD} and \hssusy~\cite{HSSUSY}.

The numerical impact of the various corrections computed in this paper
turns out to be small, typically below one GeV in regions of the MSSM
parameter space where the prediction for the Higgs mass is within a
few GeV from the observed value. We stress that this is in fact a
desirable feature of the EFT calculation of the Higgs mass: while the
logarithmically enhanced corrections are accounted for by the
evolution of the parameters between the matching scale and the EW
scale, and high-precision calculations at the EW scale can be borrowed
from the SM, the small impact of the two-loop corrections computed at
the matching scale suggests that the ``SUSY uncertainty'' associated
to uncomputed higher-order terms should be well under control.
In principle, the advantages of an EFT approach are less clear-cut
when there is only a moderate separation between the SUSY scale and
the EW scale, so that the omission of $\ovms$ effects in the
calculation of the Higgs mass is not warranted. However, our study of
the dimension-six operators suggests that the naive estimate of the
theoretical uncertainty associated to missing $\ovms$ effects (or
``EFT uncertainty'') implemented in the code \SHD\ is indeed
sufficiently conservative in the relevant regions of the MSSM
parameter space.  The EFT approach also becomes more complicated when
some of the new particles are much lighter than the rest. For example,
while our results for the two-loop corrections to the quartic Higgs
coupling can be directly applied to the standard split-SUSY scenario
by taking the limit of vanishing gluino and higgsino masses, scenarios
in which both Higgs doublets are light require a dedicated
calculation, in which the effective theory valid below the SUSY scale
is a THDM (see, e.g., ref.~\cite{Lee:2015uza}).

Finally, we recall that the accuracy of the measurement of the Higgs
mass at the LHC has already reached the level of a few hundred MeV --
i.e., comparable to the effects of the corrections discussed in this
paper -- and will improve further when more data become available. If
SUSY shows up at last, the mass and the couplings of the SM-like Higgs
boson will serve as precision observables to constrain MSSM parameters
that might not be directly accessible by experiment, especially in
scenarios where some of the superparticle masses are in the multi-TeV
range. To this purpose, the accuracy of the theoretical predictions
will have to match the experimental one, making a full inclusion of
two-loop effects in the Higgs-mass calculation unavoidable. Our
results should be regarded as necessary steps in that direction.

\section*{Acknowledgments}

We thank A.~Falkowski, M.D.~Goodsell, the authors of \FH\ and those of
{\tt FlexibleEFTHiggs} for useful discussions. The work of E.~B.~is
supported by the Collaborative Research Center SFB676 of the Deutsche
Forschungsgemeinschaft (DFG), ``Particles, Strings and the Early
Universe''. The work of P.~S.~is supported in part by French state
funds managed by the Agence Nationale de la Recherche (ANR), in the
context of the LABEX ILP (ANR-11-IDEX-0004-02, ANR-10-LABX-63) and of
the grant ``HiggsAutomator'' (ANR-15-CE31-0002). P.~S.~also
acknowledges support by the Research Executive Agency (REA) of the
European Commission under the Initial Training Network ``HiggsTools''
(PITN-GA-2012-316704) and by the European Research Council (ERC) under
the Advanced Grant ``Higgs@LHC'' (ERC-2012-ADG\_20120216-321133).

\vfill
\newpage

\begin{Appendix}

\section*{Appendix}

We present here the one-loop scalar contributions to the matching
condition for the quartic Higgs coupling, including all terms
controlled by third-family Yukawa couplings:
\bea
(4\pi)^2\,\Delta \lambda^{1\ell,\,\phi} &=& 
N_c\, \hat g_t^2 \left[\hat g_t^2 
\,+\, \frac{1}{2} \left(g_2^2-\frac{g_1^2}5\right) 
\cos  2 \beta  \right] \ln \frac{m_{Q_3}^2}{Q^2} 
~+~N_c\, \hat g_t^2  \left[\hat g_t^2 + \frac{2}{5}\, g_1^2\,
 \cos 2 \beta \right] 
\ln \frac{m_{U_3}^2}{Q^2} \nonumber \\[2mm] 
&+& N_c\, \hat g_b^2 \left[\hat g_b^2 
\,-\, \frac{1}{2} \left(g_2^2+\frac{g_1^2}5\right) 
\cos  2 \beta  \right] \ln \frac{m_{Q_3}^2}{Q^2} 
~+~N_c\, \hat g_b^2  \left[\hat g_b^2 - \frac{g_1^2}{5}\,\,
 \cos 2 \beta \right] 
\ln \frac{m_{D_3}^2}{Q^2} \nonumber \\[2mm]
&+& \hat g_\tau^2 \left[\hat g_\tau^2 
\,-\, \frac{1}{2} \left(g_2^2 - \frac35\,g_1^2 \right) 
\cos  2 \beta  \right] \ln \frac{m_{L_3}^2}{Q^2} 
~+~ \hat g_\tau^2  \left[\hat g_\tau^2 - \frac35\, g_1^2\,\,
 \cos 2 \beta \right] 
\ln \frac{m_{E_3}^2}{Q^2} \nonumber \\[2mm]
    &+&\frac{ \cos^2 2 \beta}{300}\, \sum_{i=1}^3\, 
\bigg[\,N_c\left(g_1^4+25\, g_2^4\right) \,
\ln \frac{m_{Q_i}^2}{Q^2}  \,+\,8\,N_c\, g_1^4 \,\ln \frac{m_{U_i}^2}{Q^2} 
\,+\,2\,N_c\, g_1^4 \,\ln \frac{m_{D_i}^2}{Q^2} \nonumber \\
   && ~~~~~~~~~~~~~~~~~~~~~
+\, \left(9 \,g_1^4+25\, g_2^4\right)\, \ln \frac{m_{L_i}^2}{Q^2}   
\,+\,18 \,g_1^4 \,\ln \frac{m_{E_i}^2}{Q^2}
 \bigg]  \nonumber \\[2mm]
 &+&\frac{1}{4800}\,  \bigg[261\, g_1^4+630\, g_1^2 g_2^2  +1325\,
 g_2^4  - 4 \,\cos 4 \beta  \left(9\, g_1^4+90\, g_1^2 g_2^2+175\, g_2^4\right) 
\nonumber \\
&&~~~~~~~~~~~ -9\, \cos 8\beta  \left(3\, g_1^2+5\, g_2^2\right)^2 \bigg] 
\ln \frac{\ma^2}{Q^2} -\frac3{16} \left(\frac35 g_1^2 + g_2^2\right)^2 
\sin ^2 4 \beta \nonumber \\[2mm]
&+&
\sum_{f=t,b,\tau}~\hat g_f^2 \, N^f_c\,  \wt X_f \left\{~
2\,\hat g_f^2 \, \left[\wt F_1\left(x_f\right)
  -\frac{\wt X_f}{12} \,\wt F_2\left(x_f\right)\right]
\right.  \nonumber \\
&& ~~~~~~~~~~~~~~~~~~~~~~~~+ \frac{\cos 2\beta}{4} \,
\left[\,\frac{9}{10} \, g_1^2 \,Q_f\, \wt F_3 \left(x_f\right)\right. 
\nonumber \\
&&~~~~~~~~~~~~~~~~~~~~~~~~~~~~~~~~~~~~~~~
\left. + \left( 2\, g_2^2 \,T^3_{f_\smallL} 
+ \frac35\,g_1^2\,(2\,T^3_{f_\smallL} -\frac32 \,Q_f )\right)\wt 
F_4 \left(x_f\right) \right]\nonumber \\
&&
\left. \phantom{\frac{\wt X_f}{12}}
~~~~~~~~~~~~~~~~~~~-~\frac{\cos^2 2\beta}{12} \, 
\left( \frac35 g_1^2 +g_2^2 \right) \wt F_5\left(x_f \right)~\right\}~,
\label{fulloneloop}
\eea
where the compact notation used in the sum over the sfermion species
$f=t,b,\tau$ is described after eq.~(\ref{oneloop}), and all loop
functions $\wt F_i$ are defined in appendix A of
ref.~\cite{Bagnaschi:2014rsa}. In addition, $Q_f$ is the electric
charge and $T^3_{f_\smallL}$ is the third component of the weak
isospin of the ``left'' sfermion of each species. We recall that
eq.~(\ref{fulloneloop}) assumes that the tree-level part of the
matching condition for $\lambda\,$, see eq.~(\ref{looplam}), be
expressed in terms of the EW gauge couplings of the SM and of an angle
$\beta$ defined as in section~2.2 of ref.~\cite{Bagnaschi:2014rsa}.
We also remark that the third-family Yukawa couplings $\hat g_f$
entering eq.~(\ref{fulloneloop}) are the MSSM ones. As discussed in
section~\ref{sec:lambda2loop}, our choice of using instead the top and
tau Yukawa couplings of the SM (denoted as $g_t$ and $g_\tau$) in the
one-loop part of the threshold correction to the quartic Higgs
coupling induces shifts in the two-loop part of the correction, see
eq.~(\ref{dshiftsq}).  Finally, we note that eq.~(\ref{fulloneloop})
differs from eq.~(11) of ref.~\cite{Vega:2015fna} by the presence of
the terms in the second and third lines.

\end{Appendix}

\vfill
\newpage

\bibliographystyle{utphys}
\bibliography{BPVS}

\end{document}